\documentclass[sigconf, nonacm]{acmart}
\pdfoutput=1
\makeatletter
\def\@ACM@checkaffil{
    \if@ACM@instpresent\else
    \ClassWarningNoLine{\@classname}{No institution present for an affiliation}%
    \fi
    \if@ACM@citypresent\else
    \ClassWarningNoLine{\@classname}{No city present for an affiliation}%
    \fi
    \if@ACM@countrypresent\else
        \ClassWarningNoLine{\@classname}{No country present for an affiliation}%
    \fi
}
\makeatother

\usepackage{amsmath,amsfonts}
\usepackage{algorithmic}
\usepackage{tikz}
\usepackage{graphicx}
\usepackage{textcomp}
\usepackage{xcolor}
\usepackage{booktabs}
\usepackage{pifont}
\usepackage{siunitx}
\usepackage{amsmath}
\usepackage{algorithm2e}
\usepackage{multirow}
\usepackage{fancyhdr}
\usepackage{float}
\usepackage{setspace}
\usepackage{listings}
\usepackage{balance}
\usepackage{wrapfig}
\usepackage{marginnote}
\usepackage{caption}
\usepackage{boldline}
\usepackage{colortbl}
\usepackage{floatflt}
\usepackage{hyperref}
\usepackage{eso-pic}
\usepackage{titlesec}
\usepackage{noindentafter}
\usepackage[htt]{hyphenat}
\usepackage{tablefootnote}

\sloppypar

\newcommand{\secref}[1]{§\ref{#1}}

\definecolor{whitesmoke}{rgb}{0.96, 0.96, 0.96}

\usepackage{calc}
\newcommand{\dingOne}{\ding{182}}
\newcommand{\dingTwo}{\ding{183}}
\newcommand{\dingThree}{\ding{184}}
\newcommand{\dingFour}{\ding{185}}
\newcommand{\dingFive}{\ding{186}}
\newcommand{\dingSix}{\ding{187}}
\newcommand{\dingSeven}{\ding{188}}

\newcommand\encircle[1]{%
  \tikz[baseline=(X.base)] 
    \node (X) [draw, shape=circle, inner sep=0] {\footnotesize#1};}

\newcommand\encirclenum[1]{%
  \tikz[baseline=(X.base)] 
    \node (X) [draw, shape=circle, inner sep=0] {\small#1};}

\newcounter{obs}
\newcommand\observation[1]{%
   \refstepcounter{obs}
  \vspace{0.3em}
  \noindent
  \begin{tabular}{|p{0.95\linewidth}|}
       \hline
       \textbf{{Obsv. \theobs}.} {{#1}}\\
       \hline 
  \end{tabular}
  \vspace{-0.3em}
}

\newcounter{take}
\newcounter{tkw}
\newcommand\takeawaybox[1]{%
   \refstepcounter{tkw}
   \vspace{0.3em}

  \noindent
  \begin{tabular}{V{3}p{0.95\linewidth}V{3}}
       \hlineB{3}
       \rowcolor{whitesmoke}
       \textbf{{Takeaway \thetkw}.} {{#1}}\\
       \hlineB{3}
  \end{tabular}
}

\newcommand{\shellcmd}[1]{\\\indent\indent\texttt{\footnotesize\$ #1}}
\newcommand{\shellcommand}[1]{\\\indent\indent\texttt{\footnotesize #1}}
\newcommand{\mytexttilde}{\raisebox{0.5ex}{\texttildelow}}

\definecolor{st1}{rgb}{0.580, 0.050, 0.211}
\definecolor{st2}{rgb}{0.007, 0.520, 0.867}
\definecolor{jl}{rgb}{1.0, 0.2, 0.8}
\definecolor{moegi}{rgb}{0.357, 0.537, 0.188}

\definecolor{gfored}{rgb}{0.580, 0.050, 0.211}
\definecolor{niceorange}{rgb}{0.86, 0.34, 0.12}
\definecolor{ao}{rgb}{0.007, 0.520, 0.867}
\definecolor{yt}{rgb}{0.875, 0.568, 1.000}
\definecolor{moegi}{rgb}{0.357, 0.537, 0.188}
\definecolor{jl}{rgb}{1.0, 0.2, 0.8}
\definecolor{brown(web)}{rgb}{0.65, 0.16, 0.16}
\definecolor{bisque}{rgb}{1.0, 0.89, 0.77}
\definecolor{lightbisque}{rgb}{1.0, 0.95, 0.85}
\definecolor{CadetBlue}{rgb}{0.37, 0.62, 0.63}
\definecolor{whitesmoke}{rgb}{0.96, 0.96, 0.96}
\definecolor{beautifulgreen}{rgb}{0.13, 0.55, 0.13}
\definecolor{teal}{rgb}{0.0, 0.5, 0.5}

\newif\ifcamerareadydraft
\camerareadydraftfalse

\ifcamerareadydraft
    \newcommand{\srm}[1]{{#1}}
    \newcommand{\sro}[1]{{#1}}
    \newcommand{\sr}[1]{{#1}}

    \newcommand{\hluo}[1]{{#1}}
    \newcommand{\om}[1]{\textcolor{red}{#1}}
    \newcommand{\srmf}[1]{\textcolor{cyan}{#1}}
    \newcommand{\srof}[1]{\textcolor{red}{#1}}
    \newcommand{\srf}[1]{\textcolor{blue}{#1}}
    \newcommand{\sraf}[1]{\textcolor{blue}{#1}}
    
    \newcommand{\hluof}[1]{\textcolor{blue}{#1}}
\else
    \newcommand{\srm}[1]{{#1}}
    \newcommand{\sro}[1]{{#1}}
    \newcommand{\sr}[1]{{#1}}

    \newcommand{\hluo}[1]{{#1}}
    \newcommand{\om}[1]{{#1}}
    \newcommand{\srmf}[1]{{#1}}
    \newcommand{\srof}[1]{{#1}}
    \newcommand{\srf}[1]{{#1}}
    \newcommand{\sraf}[1]{{#1}}
    
    \newcommand{\hluof}[1]{{#1}}
\fi

\AtBeginDocument{
  \providecommand\BibTeX{{
    \normalfont B\kern-0.5em{\scshape i\kern-0.25em b}\kern-0.8em\TeX}}}

\author{
{Steve Rhyner$^{1}$}\quad\hspace{0.5em}
{Haocong Luo$^{1}$}\quad\hspace{0.5em}
{Juan Gómez-Luna$^{2}$}\quad\hspace{0.5em}
{Mohammad Sadrosadati$^{1}$}\\
{Jiawei Jiang$^{3}$}\quad
{Ataberk Olgun$^{1}$}\quad
{Harshita Gupta$^{1}$}\quad
{Ce Zhang$^{4}$}\quad
{Onur Mutlu$^{1}$}\vspace{0.7em}\\
{\small$^1$ETH Zurich}
\quad
{\small$^2$NVIDIA}
\quad
{\small$^3$Wuhan University}
\quad
{\small$^4$University of Chicago}
}

\renewcommand{\shortauthors}{Rhyner, et al.}

\renewcommand{\authors}{Steve Rhyner, Haocong Luo, Juan Gómez-Luna, Mohammad Sadrosadati, Jiawei Jiang, Ataberk Olgun, Harshita Gupta, Ce Zhang, Onur Mutlu}

\setcopyright{none}
\begin{document}

\title{PIM-Opt: Demystifying Distributed Optimization Algorithms\\on a Real-World Processing-In-Memory System}

\begin{abstract}

Modern \emph{Machine Learning} (ML) training on large-scale datasets is a very time-consuming workload. It relies on the optimization algorithm \emph{Stochastic Gradient Descent} (SGD) due to its effectiveness, simplicity, and generalization performance (i.e., test performance on unseen data). Processor-centric architectures (e.g., CPUs, GPUs) commonly used for modern ML training workloads based on SGD are bottlenecked by data movement between the processor and memory units due to the poor data locality in accessing large training datasets. As a result, processor-centric architectures suffer from low performance and high energy consumption while executing ML training workloads. \emph{Processing-In-Memory} (PIM) is a promising solution to alleviate the data movement bottleneck by placing the computation mechanisms inside or near memory. Several prior works propose PIM techniques to accelerate ML training; however, prior works either do not consider \emph{real-world} PIM systems or evaluate algorithms that are not widely used in modern ML training.

Our goal is to understand the capabilities and characteristics of popular distributed SGD algorithms on \emph{real-world} PIM systems to accelerate data-intensive ML training workloads. To this end, we 1) implement several representative centralized parallel SGD algorithms, i.e., based on a central node responsible for synchronization and orchestration, on the \emph{real-world} general-purpose UPMEM PIM system, 2) rigorously evaluate these algorithms for ML training on large-scale datasets in terms of performance, accuracy, and scalability, 3) compare to conventional CPU and GPU baselines, and 4) discuss implications for future PIM hardware. We highlight the need for a shift to an algorithm-hardware codesign to enable decentralized parallel SGD algorithms in real-world PIM systems, which significantly reduces the communication cost and improves scalability.

Our results demonstrate three major findings: 1) The general-purpose UPMEM PIM system can be a viable alternative to state-of-the-art CPUs and GPUs for many memory-bound ML training workloads, especially when operations and datatypes are natively supported by PIM hardware, 2) it is important to carefully \emph{choose} the optimization algorithms that best fit PIM, and 3) the UPMEM PIM system does \emph{not} scale approximately linearly with the number of nodes for many data-intensive ML training workloads. We open source all our code to facilitate future research at \url{https://github.com/CMU-SAFARI/PIM-Opt}.

\end{abstract}

\maketitle
\thispagestyle{firstpage}
\fancypagestyle{firstpage}
{
    \fancyhead{}
    \begin{tikzpicture}[remember picture,overlay]
    \node [xshift=191mm,yshift=-10mm]
    at (current page.north west) {\href{https://www.acm.org/publications/policies/artifact-review-and-badging-current}{\includegraphics[width=1.8cm]{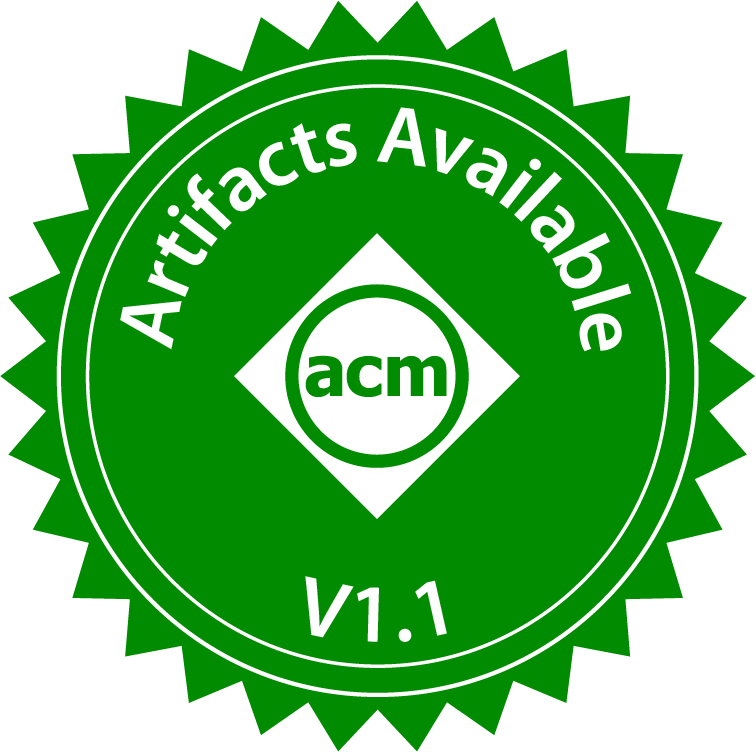}}} ;
    \end{tikzpicture}

  \renewcommand{\headrulewidth}{0pt}
}
\thispagestyle{firstpage}

\pagestyle{fancy}
\fancyhead{}

\section{Introduction}
\label{sec:Introduction}

\emph{Stochastic Gradient Descent} (SGD)~\cite{robbins1951stochastic,boyd2004convex} is perhaps the most important and commonly deployed optimization algorithm for modern \emph{\sr{M}achine \sr{L}earning} (ML) training~\cite{stich2018sparsified,bottou2010large,jiang2021towards,bottou2012stochastic,kingma2014adam,li2014efficient}. SGD is the main building block of most centralized and decentralized optimization algorithms that have been introduced to accommodate the continuously increasing demand for scalability and high-performance training of ML models on large-scale datasets.

Training ML models on growing datasets~\cite{villalobos2022will,wang2020survey,dunner2018snap} is a time-consuming \srm{task that demands both high computational power and memory bandwidth}~\cite{epoch2023trendsinthedollartrainingcostofmachinelearningsystems,gomezluna2021prim,gomez2021benchmarking,gomez2022benchmarking}. \sro{The low data reuse during ML training on large-scale datasets leads to poor data locality.} As a result, processor-centric architectures (e.g., CPU, GPU) commonly used by the ML community repeatedly need to move training samples between the processor and off-chip memory. This not only degrades performance~\cite{ivanov2021data} but is also a major source of the overall system's energy consumption~\cite{boroumand2018google}. This phenomenon is referred to as the data movement bottleneck \om{\cite{mutlu2020intelligent,mutlu2019processing,mutlu2022modern}}, which is \om{common} in data-intensive workloads. ML training \om{is a} prominent example \om{of such workloads}.

\emph{Processing-In-Memory} (PIM)~\cite{mutlu2019processing,mutlu2022modern,mutlu2023dac,ghose2019processing,seshadri2019dram,mutlu2019enabling} is \srm{a promising way} to alleviate the data movement bottleneck by placing the computation mechanisms inside or near memory \srm{units}. PIM, an idea proposed several decades ago~\cite{kautz1969cellular,stone1970logic}, is a \om{memory-centric} comput\om{ation} paradigm that \sro{has} recently gained traction in \sro{both} academia\om{~\cite{mutlu2020intelligent,mutlu2019processing,mutlu2022modern,mutlu2023dac,ghose2019processing,seshadri2019dram,mutlu2019enabling,gomezluna2021prim,gomez2021benchmarking,gomez2022experimental,gomez2023evaluatingispass,khan2024landscape,giannoula2022sparsep,giannoula2024accelerating,hyun2023pathfinding,hyun2024pathfinding,10158230,nider2021case,peccerillo2022survey,abumpimp2023,chen2023simplepim,khan2022cinm,chen2023uppipe,gogineni2024swiftrl,diab2023framework,lavenier2020variant,lavenier2016blast,gupta2023evaluating,jonatan2024scalability,bernhardt2023pimdb,lim2023design,baumstark2023adaptive,baumstark2023accelerating,kang2023pim,das2022implementation,Zarif_2023,kim2024optimal,wu2023pim,gao2015practical,falahati2018origami,vieira2018exploiting,sun2020one,shelor2019reconfigurable,saikia2019k,deng2018dracc,boroumand2018google,gao2017tetris,boroumand2021google,cho2020mcdram,shin2018mcdram,azarkhish2017neurostream,kwon2019tensordimm,ke2020recnmp,cordeiro2021machine,lee2021task,park2021high,park2021trim,kim2020mvid}} \sro{and industry\om{;}} some commercial \om{PIM} systems and prototypes \sro{have \om{recently} been developed by} industry\sro{~\cite{upmemwebsite2024, upmemtechpaper2022,upmemproductsheet2022,devaux2019true,kwon202125,ke2021near,lee2021hardware,lee20221ynm,niu2022184qps}}.

\srm{Several prior works \sro{explore} the effectiveness of using PIM for fundamentally improving ML training performance and energy efficiency} \om{\cite{gao2015practical,falahati2018origami,vieira2018exploiting,sun2020one,shelor2019reconfigurable,saikia2019k,liu2018processing,luo2020benchmark,sun2020energy,li20203d,imani2019floatpim,jiang2019cimat,schuiki2018scalable}}. However, none of these prior works provide a comprehensive evaluation on \emph{real-world} general-purpose PIM architectures. To our knowledge, there is only one prior work~\cite{gomez2022experimental,gomez2023evaluatingispass} on training and evaluating ML models on a \emph{real-world} PIM system using \emph{\sr{G}radient \sr{D}escent} (GD)~\cite{polyak1987introduction}\om{. U}nfortunately, GD-based algorithms are not widely used in modern \srm{ML}. \srm{SGD\sro{~\cite{robbins1951stochastic,boyd2004convex}} is a simplification of GD\om{:} in each iteration, only stochastic gradients instead of the full gradient need to be computed~\cite{bottou2012stochastic}. Since stochastic gradients are\sr{,} in general\sr{,} significantly more efficient to compute compared to full gradient\sro{s}, SGD alleviates the computational bottleneck~\cite{stich2018sparsified} \sro{of computing the full gradient by approximating the expected gradient with an unbiased estimate~\cite{goodfellow2016}}. Variants of SGD such as \sr{m}ini-\sr{b}atch SGD~\cite{li2014efficient} allow for parallelization by batching \om{(in each iteration)} the training samples whose gradients can be computed independently.} We focus on popular SGD-based algorithms due to their effectiveness, simplicity, and generalization performance \om{(i.e., test performance on unseen data)}~\cite{zhou2020towards}.

\textbf{Our goal} in this paper is to understand the capabilities and characteristics of popular distributed \srm{SGD} algorithms on \emph{real-world} PIM architectures to accelerate data-intensive ML training workloads. To do so, we implement and rigorously evaluate $12$ representative ML training workloads, commonly used in the ML community, on \om{the} \emph{real-world} UPMEM PIM architecture. We choose \om{the} \srm{general-purpose} UPMEM PIM system for our study because it is \srm{commercially \sro{available}}~\cite{upmemwebsite2024,upmemtechpaper2022,upmemproductsheet2022}. First, we implement and investigate all combinations of 1) three \om{centralized parallel} \sro{SGD} algorithms that specifically take into account the close resemblance of \srm{the UPMEM PIM system} to a distributed system~\cite{chen2023simplepim}, \emph{\sr{m}ini-\sr{b}atch Stochastic Gradient Descent with Model Averaging} (MA-SGD)~\cite{zinkevich2010parallelized,mcdonald2010distributed}, \emph{\sr{m}ini-\sr{b}atch Stochastic Gradient Descent Gradient Averaging} (GA-SGD)~\cite{dekel2012optimal,li2014communication}, and the \emph{communication-efficient} distributed \emph{Alternating Direction Method of Multipliers} (ADMM) algorithm~\cite{MAL-016_Boyd_ADMM}, 2) two popular and representative linear binary classification models, \emph{Logistic Regression} (LR)\sro{~\cite{goodfellow2016}} and \emph{Support Vector Machine} (SVM)\sro{~\cite{boser1992training,cortes1995support,goodfellow2016}}, and 3) two large-scale datasets, YFCC100M-HNfc6 and Criteo~\cite{amato2016yfcc100m,criteo1tb2014}. Second, we rigorously evaluate all combinations of these algorithms, models, and datasets in terms of performance, accuracy, and scalability. Third, we compare the training speed and test set inference accuracy of \srm{the UPMEM PIM system} to state-of-the-art CPU (2x AMD EPYC 7742 CPU 64-core processor\sro{~\cite{amdepyc2019}}) and GPU (NVIDIA A100\sro{~\cite{nvidia2020a100}}) baselines. Fourth, we discuss implications for future PIM hardware and \om{highlight} the need \om{for a} shift to an algorithm-hardware codesign perspective to accommodate decentralized optimization algorithms \sro{on \emph{real-world} PIM systems by supporting direct communication across PIM nodes}. 

\sro{\bf{Our results}} \om{demonstrate} three major findings: 1) \srm{\om{t}he UPMEM PIM system} can be a viable alternative to state-of-the-art CPUs and GPUs for many data-intensive ML training workloads when operations and datatypes are natively supported by PIM hardware\om{.} \om{For instance, for the YFCC100M-HNfc6 (Criteo) dataset, training SVM with GA-SGD on PIM \sr{is} $1.94$x faster ($2.43$x slower) compared to the CPU baseline, and $3.19$x ($10.65$x) faster compared to \sr{mini-batch} SGD on the GPU architecture while achieving similar accuracy.} 2) \sro{It is} importan\sro{t} \sro{to} carefully \sro{\emph{choose}} the optimization algorithm that best fits PIM. \om{\om{For example}, for the YFCC100M-HNfc6 (Criteo) dataset, training SVM with the ADMM algorithm using PIM, we observe speedups of $3.19$x ($31.82$x) compared to GA-SGD at the cost of a \sro{small} reduction\sro{, i.e., $1.007$x ($1.014$x), in} test accuracy (AUC score\sr{~\cite{huang2005using}}; see~\secref{sec:datasets} for more details).} 3) \sro{The UPMEM PIM system exhibits scalability challenges for many ML training workloads in terms of \emph{statistical efficiency}, i.e., how many steps are needed until convergence~\cite{zhang2014dimmwitted}}. \om{\om{For instance}, in our strong scaling (see~\secref{sec:evaluation} for more details) experiments of the YFCC100M-HNfc6 (Criteo) dataset, training LR with ADMM using PIM, we observe speedups of $7.43$x ($3.85$x) while the achieved test accuracy (AUC score) decreases from $95.46$\% ($0.74$) to $92.17$\% ($0.718$), as we scale the number of nodes from $256$ to $2048$.} \sro{This reduction in accuracy is due to the fact that more nodes increase staleness as each node \om{uses} its own local model before synchron\om{izing} with the central node.}\footnote{\om{For a theoretical analysis of this phenomenon of how the number of nodes affects the convergence rate, we refer the reader to~\cite{zhou2017convergence}.}}

This paper makes the following key contributions:
\begin{itemize}
    \item To our knowledge, \sro{this paper is} the first to implement, analyze, and train linear ML models \om{on two large-scale datasets} using \emph{realistic} and \emph{communication-efficient} distributed \sr{SGD} algorithms on a \emph{real-world} PIM \srm{system} \sro{(i.e., UPMEM)}.
    \item We present the design space covering the design choices of various algorithms, models, and workloads for ML training on \om{a state-of-the-art} \sro{\emph{real-world}} PIM \sro{system}.
    \item We demonstrate scalability challenges \srm{of \om{the} UPMEM PIM system} in terms of \emph{statistical efficiency}. \sro{We} discuss implications for hardware design to accommodate decentralized optimization algorithms and \om{highlight} the need \om{for a} shift towards an algorithm-hardware codesign in the context of ML training using PIM.
    \item We open source all our code to facilitate future research \sro{at \url{https://github.com/CMU-SAFARI/PIM-Opt}}.
\end{itemize}
\section{Background \& Motivation}

\noindent We provide a brief introduction \srm{to} \srmf{linear \srm{m}odels,} \srmf{\emph{Machine Learning} (ML)} training, regularization, and \srm{a}lgorithms (\secref{sec:model_algo_dataset}). We \sro{describe} \om{the} UPMEM \emph{Processing-In-Memory} (PIM) system (\secref{sec:UPMEM_PIM_arch}), the first \emph{real-world} general-purpose PIM hardware architecture \srm{that we \sro{perform} ML training on}. \srm{There are a number of works exploring a variety of approaches on PIM~\cite{chang2017understanding,seshadri2013rowclone,seshadri2015fast,chang2016low,seshadri2016buddy,seshadri2017ambit,hajinazar2021simdram,seshadri2017simple,seshadri2016simple,boroumand2018google,boroumand2021google,cali2020genasm,hashemi2016accelerating,hashemi2016continuous,ahn2015scalable,ahn2015pim,boroumand2020practical,zhu2013accelerating,pugsley2014ndc,zhang2014top,farmahini2015nda,hsieh2016transparent,pattnaik2016scheduling,akin2015data,hsieh2016accelerating,lee2015bssync,gao2016hrl,chi2016prime,gu2016biscuit,kim2016neurocube,asghari2016chameleon,boroumand2016lazypim,liu2017concurrent,hassan2015near,nai2017graphpim,kim2017grim,fernandez2020natsa,singh2019napel,herruzo2021enabling,boroumand2021polynesia,giannoula2021syncron,besta2021sisa,asgari2021fafnir,denzler2023casper,oliveira2023dappa,sun2021abc,lee2022improving,dai2022dimmining,boroumand2019conda,drumond2017mondrian,dai2018graphh,zhuo2019graphq,oliveira2022heterogeneous,oliveira2024mimdram,yuksel2024functionally,park2022flash,mansouri2022genstore,ghiasi2024megis,mao2022genpip,gao2019computedram,fernandez2024matsa,shahroodi2023swordfish,olgun2022pidram,ghiasi2022alp,oliveira2021damov,singh2020nero}.} For general PIM background and discussion of many works in the field, we refer the reader to\sro{~\cite{mutlu2022modern,mutlu2019processing,ghose2019processing,mutlu2023dac}}. We conclude this section with our \srm{m}otivation (\secref{sec:motivation}).

\subsection{Models, ML Training, Regularization \& Algorithms}
\label{sec:model_algo_dataset}

\sloppy

\noindent\textbf{Models.} \srm{Two of the most commonly trained} linear binary classification models for convex optimization tasks \srm{are}: 1) \emph{Logistic Regression} (LR) with Binary Cross Entropy Loss (BCE)~\cite{goodfellow2016}, and 2) \emph{Support Vector Machines} (SVM) with Hinge Loss~\cite{boser1992training,cortes1995support,goodfellow2016}. Each model consists of a linear layer and a bias.

\noindent\textbf{ML Training.} The goal of \srmf{\emph{Machine Learning} (ML)} training \srm{is to find an optimal ML model
\begin{equation}
    \label{eq:equation_1}
    w^* = \underset{w \in \mathbb{R}^d}{\text{argmin }} L(w) \text{ where } L(w) = \frac{1}{n} \sum_{i=1}^{n} l(x_i,y_i,w)
\end{equation}
over a training dataset $\mathcal{D} = \{ (x_i, y_i) \}_{i = 1}^{n} $~\cite{jiang2021towards,li2014efficient,goodfellow2016,hastie2009elements}}. Here, $x_i\in \mathbb{R}^{d}$ is referred to as feature vector, $y_i\in\mathbb{R}$ as label of the $i^{\text{th}}$ \sr{training} sample, \srm{$n$ denotes the cardinality of $\mathcal{D}$}, and $l(x_i,y_i,w)$ is a loss function~\cite{jiang2021towards,li2014efficient}. \srmf{For binary classification tasks, for LR, it is common to assign labels $y_i\in \{0,1\}$ to denote the membership of the $i^{\text{th}}$ training sample to one of the two classes. In contrast, for SVM, the labels are $y_i\in \{-1,1\}$.}

\noindent \srm{\textbf{Regularization.} It is common to add a regularizer $r(w)$ to prevent overfitting on the training dataset and to control model complexity. The objective function is obtained by setting 
\begin{equation}
    l(x_i,y_i,w) = l'(x_i,y_i,w) + \lambda r(w)
\end{equation}
where \om{$l'(x,y,w)$ is a loss function} and $\lambda$ is the regularization parameter~\cite{li2014efficient}. The regularization strategy of defining the regularizer as $r(w)=\frac{1}{2}||w||_2^2$ is referred to as $L^2$ regularization~\cite{goodfellow2016}. Another popular approach is to set $r(w)=||w||_1$, i.e., the sum of the absolute values of the model parameters, known as $L^1$ regularization~\cite{goodfellow2016}.}

\noindent\textbf{Algorithms.} \srm{\emph{Stochastic Gradient Descent} (SGD)~\cite{robbins1951stochastic,boyd2004convex} is perhaps the most important and commonly deployed optimization algorithm for modern \emph{Machine Learning} (ML) training~\cite{stich2018sparsified,bottou2010large,jiang2021towards,bottou2012stochastic,kingma2014adam,li2014efficient}. In each iteration, SGD computes a stochastic gradient and updates the model~\cite{ruder2016overview}.} \srm{SGD is a simplification of GD\sr{~\cite{polyak1987introduction}}\om{:} in each iteration, only stochastic gradients instead of the full gradient need to be computed~\cite{bottou2012stochastic}. Since stochastic gradients are\srm{, in general,} significantly more efficient to compute compared to full gradient\sro{s}, SGD alleviates the computational bottleneck~\cite{stich2018sparsified} \sro{of computing the full gradient by approximating the expected gradient with an unbiased estimate~\cite{goodfellow2016}}. Variants of SGD such as \sr{m}ini-\sr{b}atch SGD~\cite{li2014efficient} allow for parallelization by batching the training samples in each iteration whose gradients can be computed independently.} 

\srm{We provide background on} three \srm{widely used} centralized optimization algorithms for training both LR and SVM: 1) \emph{\sr{m}ini-\sr{b}atch Stochastic Gradient Descent with Model Averaging} (MA-SGD)~\cite{zinkevich2010parallelized,mcdonald2010distributed}, 2) \emph{\sr{m}ini-\sr{b}atch Stochastic Gradient Descent Gradient Averaging} (GA-SGD)~\cite{dekel2012optimal,li2014communication}, and 3) \emph{\sr{d}istributed Alternating Direction Method of Multipliers} (ADMM)~\cite{MAL-016_Boyd_ADMM}. These algorithms are based on a parameter server\srm{, i.e., a central node responsible for synchronization and updating the global model, and several workers} among which the training dataset is evenly partitioned. 

\om{In} MA-SGD, every worker trains a local model using the \om{mini-batch SGD} \om{optimization algorithm} independently and in parallel. Each worker processes several mini-batches, updating its local model before synchronization on the parameter server where the models are averaged. Then, the averaged model, i.e., the global model, is broadcast back to the workers, and each worker continues training with \sr{mini-batch} SGD starting from the global model. There exists a \emph{one-shot averaging}~\cite{zinkevich2010parallelized,mcdonald2010distributed} variant of the MA-SGD algorithm, where models are averaged after each worker has processed its entire partition of the training data. Although one-shot averaging reduces communication, it has been shown that increasing the model averaging frequency \om{leads to} a higher convergence rate~\cite{zhang2016parallel,yu2019parallel}. 

In contrast, GA-SGD distributes each batch among all workers. Each worker runs \sr{mini-batch} SGD \sr{independently and in parallel}, computes the gradients for a fraction of the batch, and communicates the gradient with the parameter server \sr{after \emph{every} iteration}, where the gradients are averaged and the global model is updated. Subsequently, the global model is communicated with the workers, and the next batch is processed. For \sr{both} GA-SGD and MA-SGD, \sr{it is common to refer} to one \emph{global epoch} once the whole training dataset has been processed.

The distributed ADMM algorithm follows a \emph{decomposition-coordination} procedure, dividing a convex optimization problem into smaller local subproblems distributed among workers~\cite{jiang2021towards,MAL-016_Boyd_ADMM}. Each worker solves its subproblem, e.g., with \sr{mini-batch} SGD until convergence. Next, the local models are communicated with the parameter server, where the global model and auxiliary variables that help lead the workers to a consensus are computed. Each worker continues training with its local model after the synchronization step. For ADMM, \sr{it is common to refer} to one \emph{global epoch} once the synchronization step on the parameter server has been completed.

\subsection{UPMEM PIM System Architecture}
\label{sec:UPMEM_PIM_arch}

\begin{figure}[h]
    \centering
    \includegraphics[width=0.89\linewidth]{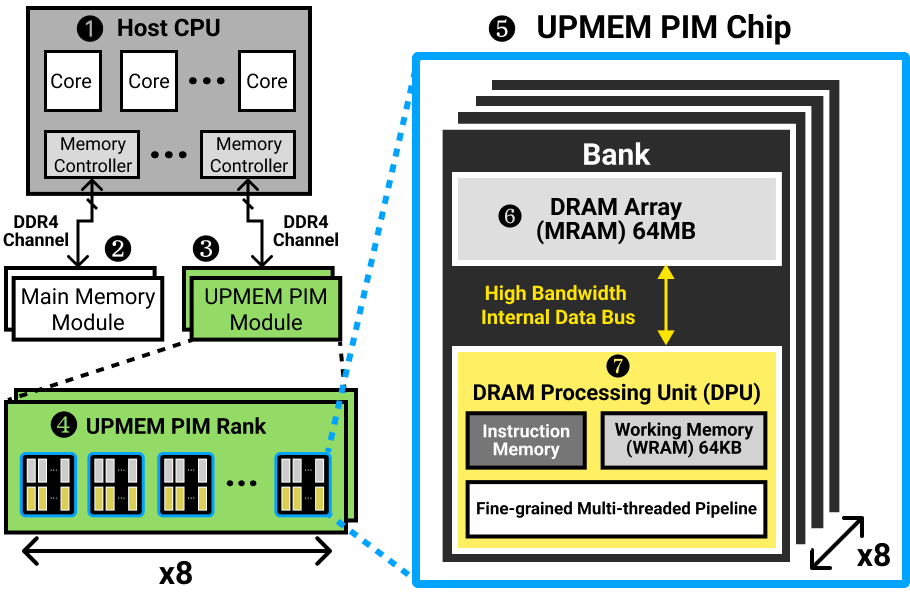}
    \caption{\om{High-level system organization of an UPMEM PIM-enabled system and the hardware architecture of an UPMEM PIM chip.}}
    \vspace{0.3em}
    \label{fig:background_PIM_figure}
\end{figure}

Fig.~\ref{fig:background_PIM_figure} \sr{shows} the high-level system organization of an UPMEM PIM-enabled system\om{~\cite{gomez2021benchmarking,gomezluna2021prim,gomez2022benchmarking}} and the hardware architecture of an UPMEM PIM chip. The system consists of a regular host CPU \dingOne, conventional main memory modules \dingTwo, and UPMEM PIM memory modules \dingThree. Each UPMEM PIM memory module contains two \emph{ranks} \dingFour. Each rank has 8 UPMEM PIM \emph{chips} \dingFive. Inside each chip, there are 8 \emph{banks}. Each bank contains 1) a 64MB DRAM array called MRAM \dingSix, and 2) a \emph{general-purpose} \emph{DRAM Processing Unit} (DPU) \dingSeven.
\vspace{0.3em}

The MRAM implements a standard JEDEC DDR4 DRAM interface that can be accessed by the host CPU. The DPU has an SRAM Instruction Memory, a 64KB SRAM Working Memory (WRAM), and an in-order fine-grained multi-threaded pipeline with 11 stages and support\srm{s} 24 hardware threads. It \srm{implements} a 32-bit RISC-based \sr{I}nstruction \sr{S}et \sr{A}rchitecture (ISA) with native support for 32-bit integer additions/subtractions and 8-bit integer multiplications. Other more complex \srm{arithmetic} operations (e.g., integer divisions and floating-point operations) are emulated through software. The DPU does not have a cache but uses the WRAM as a scratchpad memory~\cite{upmemproductsheet2022,upmemtechpaper2022}.

Each DPU has \emph{exclusive} access to its MRAM (with respect to other DPUs) through a high-bandwidth (up to 0.7GB/s \om{per DPU}) internal data bus ~\cite{gomezluna2021prim, gomez2022experimental}. There are no direct communication channels among DPUs within an UPMEM PIM chip. All inter-DPU communications are done through the host CPU (i.e., the \sr{host} CPU first gathers data from the DPUs' MRAM into the system's main memory \srm{and then distributes the data from the} main memory to the DPUs' MRAM).

\noindent\textbf{PIM Programming and Execution Model.} DPU programs are written in the C programming language with the UPMEM SDK~\cite{upmem_sdk} and runtime libraries. The execution model of \sr{a} DPU is based on the \emph{Single-Program Multiple-Data} (SPMD) paradigm. Each DPU runs multiple (up to 24) software threads, called \emph{tasklets}, which execute the same code but \srm{operate} on different data. Each tasklet has its own control flow, independent from other tasklets. Tasklets are assigned to DPUs \emph{statically} by the programmer during \emph{compile-time}. Tasklets assigned to the same DPU share MRAM and WRAM~\cite{gomezluna2021prim, gomez2022experimental}. 
\clearpage

\subsection{Motivation}
\label{sec:motivation}

\sr{\emph{Stochastic Gradient Descent} (SGD)~\cite{robbins1951stochastic,boyd2004convex}} is one of the most important optimization algorithms and the basis of many distributed optimization algorithms. However, SGD is memory-bound~\cite{xie2017cumf_sgd,de2017understanding,kim2021gradpim,mahajan2016tabla,wang2017memory}, which poses a fundamental challenge for processor-centric architectures (e.g., CPU, GPU). SGD's memory-boundedness is attributed to \om{large training} data\om{set} size\om{,} leading to decreased cache efficiency \srm{and low data reuse of training samples during ML training\om{,} which results} in performance degradation~\cite{xie2017cumf_sgd,chin2015fast,chin2015learning}. The increasing discrepancy in performance between fast processors and slow memory units exacerbates this problem~\cite{gomez2022experimental}. PIM is \srm{a promising} way to alleviate the data movement bottleneck and is a promising \om{paradigm to efficiently execute} ML training workloads. 

There are several prior proposals on PIM acceleration for ML training~\cite{gao2015practical,falahati2018origami,vieira2018exploiting,sun2020one,shelor2019reconfigurable,saikia2019k}. However, none of these prior works provide a comprehensive evaluation on \emph{real-world} general-purpose PIM \sr{systems}. To our knowledge, there is only one prior work~\cite{gomez2022experimental,gomez2023evaluatingispass} on training and evaluating ML models on a \emph{real-world} PIM system \sr{(i.e., UPMEM)} using \emph{\sr{G}radient \sr{D}escent} (GD)~\cite{polyak1987introduction}. \om{In this work, we examine popular SGD-based algorithms due to their better effectiveness, simplicity, and generalization performance~\cite{zhou2020towards}.} We specifically \sr{address} \sr{the resemblance of} \om{the} UPMEM PIM \srm{system} \sr{to} a distributed system\sr{~\cite{chen2023simplepim}} with the host CPU as \sr{the} central node\sr{, i.e., the parameter server}.

\begin{figure}[h]
    \centering
    \includegraphics[width=\linewidth]{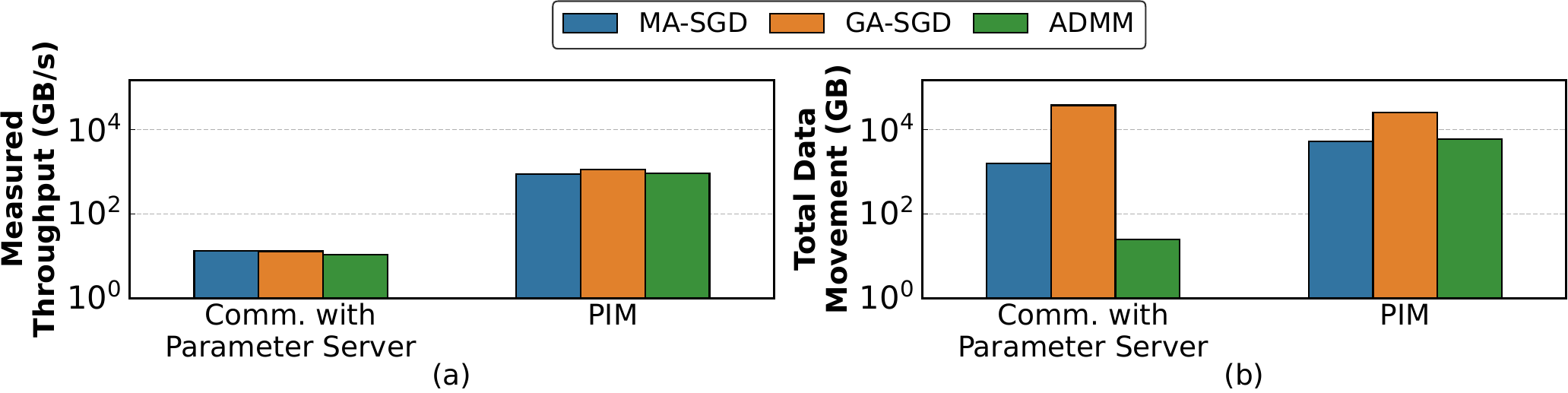}
    \caption{Per global epoch comparison of \srmf{measured throughput (a) and} \om{\srmf{total} data movement} \srmf{(b) for} all \om{studied} algorithms \om{(MA-SGD, GA-SGD, and ADMM)} \om{on the UPMEM PIM system} for the Criteo dataset.}
    \label{fig:motivational_figure}
\end{figure}

\om{We first demonstrate that it is} importan\sro{t} \sr{to} carefully \emph{choos\sro{e}} the distributed optimization algorithm that best fits \srm{the UPMEM PIM system}. \om{To do so,} we analyze key differences in \om{data movement} of distributed \sr{optimization} algorithms on \om{the} UPMEM PIM system. Fig.~\ref{fig:motivational_figure} \om{shows the per global epoch comparison of \om{data movement}} \om{for all distributed optimization algorithms we study\om{, i.e.,} MA-SGD, GA-SGD, and ADMM\om{,} using the UPMEM PIM system with $2048$ DPUs \om{training an LR model} on the Criteo dataset (see~\secref{sec:implementation} and~\secref{sec:methodology}).} \srmf{Fig.~\ref{fig:motivational_figure}(a)} shows the per global epoch \om{measured throughput} between PIM and the parameter server \srm{(\srmf{\texttt{Comm. with Parameter Server}}, i.e., the measured throughput between UPMEM PIM memory modules and the host CPU over the DDR4 channels)} and within PIM \srm{(}\srmf{\texttt{PIM}}, i.e., \srm{the internal} aggregated \om{measured throughput} between MRAM and WRAM\srm{)}. \srmf{Fig.~\ref{fig:motivational_figure}(b)} shows the per global epoch total data movement between PIM and the parameter server \sr{(\srmf{\texttt{Comm. with Parameter Server}}, i.e., the absolute amount of data exchanged between the UPMEM PIM memory modules and the host CPU over DDR4 channels)} and within PIM \sr{(\srmf{\texttt{PIM}}, i.e., the absolute amount of data transferred between MRAM and WRAM)}. For MA-SGD \srm{and} ADMM, the batch size is $2$K; for GA-SGD, it is $262$K.

We make two major observations. First, the \om{throughput} between PIM and the parameter server and within PIM is \om{very large}. For instance, for LR, we observe that the \om{throughput} within PIM for MA-SGD/GA-SGD/ADMM is $64.55$x/$88.35$x/$85.22$x higher than the \om{throughput} between PIM and the parameter server. \om{This is because the bandwidth \emph{within} PIM is much higher than the bandwidth between PIM and the parameter server.} \om{Hence, it is specifically crucial to minimize the communication between PIM and the parameter server to significantly reduce the total communication complexity.} Second, the absolute data movement between PIM and the parameter server is very high. For instance, for LR, the algorithms MA-SGD (GA-SGD) exhibit $64.01$x ($1536.14$x) higher absolute data movement for expensive communication between PIM and the parameter server per global epoch compared to ADMM. \om{This observation shows that the communication patterns imposed by MA-SGD and GA-SGD lead to a communication bottleneck on the parameter server due to the large amount of data to be transferred between PIM and the parameter server. In contrast, ADMM's \emph{efficient} communication pattern alleviates this communication bottleneck by \om{reducing} the data movement over the low-bandwidth channels between PIM and the parameter server. \om{We conclude that ADMM is a good fit for the UPMEM PIM system because it addresses the communication bottleneck on the parameter server.}}
\section{\sr{UPMEM PIM System} Implementation}
\label{sec:implementation}

\begin{figure}[h]
    \centering 
    \includegraphics[width=\linewidth]{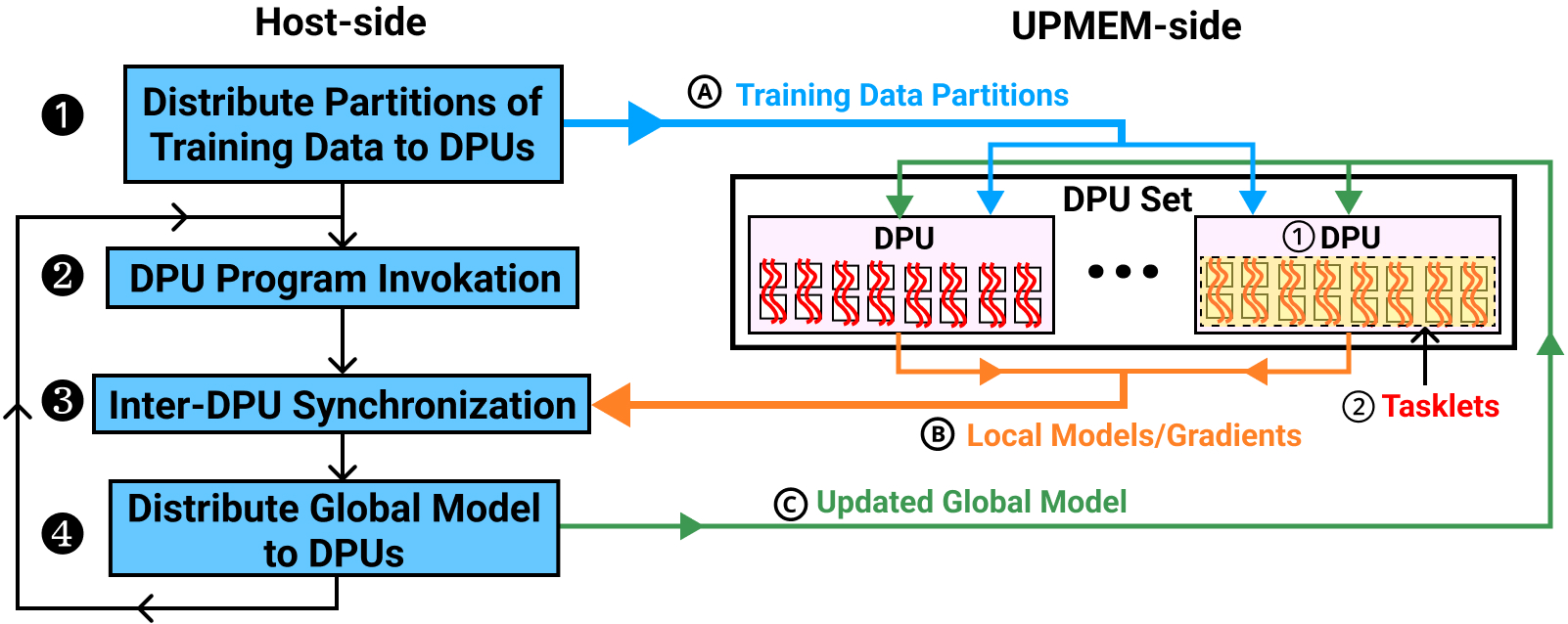}
    \caption{High-level \om{workflow for} \sr{distributed} optimization algorithms on \om{the} UPMEM PIM \sr{system}.}
    \label{fig:flowchart}
\end{figure}

\srm{\noindent Fig.~\ref{fig:flowchart} \sr{shows} the high-level \om{workflow} of training ML models \om{using distributed optimization algorithms on} the UPMEM PIM system. \om{First (\dingOne \ in Fig.~\ref{fig:flowchart}), the host CPU \emph{statically} partitions, assigns, and distributes the training data to the MRAM of the DPUs \encirclenum{1} \ (i.e., each DPU receives a partition of the whole training data \encircle{A}) and assign\srmf{s} tasklets \encirclenum{2}\ to DPUs. This training data transfer from the host to the DPUs happens \emph{only once} throughout the entire training process. Second \dingTwo, the host invokes the DPU program to run mini-batch SGD on every DPU. Third \dingThree, the host synchronizes all DPUs by collecting and aggregating the local models (MA-SGD and ADMM) or gradients (GA-SGD) \encircle{B} from the DPUs in the host's main memory to produce an updated global model \encircle{C} (see \secref{sec:model_algo_dataset}). Fourth \dingFour, the host distributes the updated global model \encircle{C} to each DPU and then invokes the DPU program \dingTwo \ again to keep on training. Steps \dingTwo, \dingThree, and \dingFour \ repeat until the training finishes (i.e., reaches a certain number of global epochs or achieves a certain level of accuracy).}}

\noindent\om{\textbf{Data Partitioning}. For both MA-SGD and ADMM, each DPU's partition consists of multiple mini-batches of the entire training data. For GA-SGD, each partition consists of a fraction of \emph{all} the mini-batches of the training data (i.e., different DPUs get different fractions of the same mini-batches).}

\noindent\om{\textbf{Synchronization.} For MA-SGD, each DPU only processes one mini-batch from its assigned training data partition and updates its local model before synchronization (i.e., model averaging) on the host. Doing so leverages the fact that increasing model averaging frequency leads to a higher convergence rate at the cost of higher communication overhead~\cite{zhang2016parallel,yu2019parallel}. For GA-SGD, each DPU computes intermediate gradients from its assigned fraction of one mini-batch before synchronization (i.e., gradient averaging) on the host. A larger batch size leads to a higher number of samples to be processed by the DPU before synchronization (i.e., less communication overhead) with the host CPU. For ADMM, each DPU processes \emph{all} assigned mini-batches and updates its local model for every mini-batch. ADMM only synchronizes the local models on the host \emph{once} after the DPUs \om{finish} processing all their assigned training data, making it attractive for distributed ML due to the low communication overhead compared to MA-SGD and GA-SGD.}

\noindent\srm{\textbf{Task Parallelism.} \om{For all distributed optimization algorithms we study (MA-SGD, GA-SGD, and ADMM), we consider every DPU as a \emph{worker}. For each DPU (worker), we use $16$ tasklets collaboratively (i.e., parallelizing the dot product and transferring data between MRAM and WRAM) to implement the mini-batch SGD optimizer }to fully utilize the multi-threaded pipeline and improve latency~\cite{gomezluna2021prim}. We evenly distribute features of the training samples and the corresponding model parameters among tasklets.}

\noindent\srm{\textbf{LUT-based Methods.} Training of LR involves computing the exponential function to evaluate the sigmoid activation function. Since \om{the} UPMEM PIM \srm{system} does not support transcendental functions, we use efficient LUT-based methods~\cite{10158230,gomez2022experimental,ferreira2022pluto} for computation. LUTs are fast~\cite{10158230,ferreira2022pluto} but incur significant storage overhead (in our case, $4$MB of MRAM per DPU). However, allocating this fraction of MRAM for the LUT is necessary to enable the evaluation of the \om{sigmoid activation} function with high precision.}

\section{Methodology}
\label{sec:methodology}

In this section, we describe the system configurations (\secref{sec:system_configuration}), \om{CPU \& GPU} \srm{b}aseline \srm{i}mplementations (\secref{sec:baseline_implementations})\sr{,} \srm{experiment} \srm{i}mplementation \srm{d}etails of \srm{our UPMEM PIM \srmf{system} \om{and} baselines} (\secref{sec:experimental_settings}), and datasets (\secref{sec:datasets}) used in this paper.

\subsection{System Configurations}
\label{sec:system_configuration}
Table~\ref{table:dpu_system-configuration} shows the system configuration of 1) the UPMEM PIM system~\sr{\cite{upmemproductsheet2022,upmemtechpaper2022}} \hluo{with 20 UPMEM PIM memory modules (2560 DPUs)}, 2) the CPU baseline system~\sr{\cite{amdepyc2019}} \hluo{with $2$\sr{x} AMD EPYC 7742 64-core CPUs (in total 128 cores)}, and 3) the GPU baseline system~\sr{\cite{nvidia2020a100}} \hluo{with an NVIDIA A100 GPU} that we \sro{perform} ML training on.

\begin{table}[h]
\centering
\caption{\sr{S}ystem \sr{C}onfigurations}
\label{table:dpu_system-configuration}
\resizebox{\linewidth}{!}{
\begin{tabular}{ll}
\hline
\hline

\multicolumn{2}{c}{\textbf{UPMEM PIM System}} \\

\hline
\textbf{Processor}                                                   & \begin{tabular}[c]{@{}l@{}} 2x Intel Xeon Silver 4215 8-core processor @ 2.50GHz\end{tabular}  \\ \hline
\textbf{Main Memory}                                                        & \begin{tabular}[c]{@{}l@{}}\SI{256}{\giga\byte} total capacity\\4$\times{}$\SI{64}{\giga\byte} DDR4 (RDIMMs)\end{tabular}  \\ \hline
\begin{tabular}[c]{@{}l@{}}\textbf{PIM-Enabled}\\\textbf{Memory}\end{tabular} & \begin{tabular}[c]{@{}l@{}} \SI{160}{\giga\byte} total capacity\\20$\times{}$\SI{8}{\giga\byte} UPMEM PIM modules,\\2560 DPUs,\\2 ranks per module, 8 chips per rank, 8 DPUs per chip\\\SI{350}{\mega\hertz} DPU clock frequency\end{tabular}   \\ \hline
\hline
\multicolumn{2}{c}{\textbf{CPU Baseline System}} \\
\hline

\textbf{Processor}                                                   & \begin{tabular}[c]{@{}l@{}} 2x AMD EPYC 7742 64-core processor @ 2.25GHz\end{tabular}  \\ \hline
\textbf{Main Memory}                                                        & \begin{tabular}[c]{@{}l@{}}\SI{1}{\tera\byte} total capacity\\32$\times{}$\SI{32}{\giga\byte} DDR4 (RDIMMs)\end{tabular}  \\ \hline
\hline
\multicolumn{2}{c}{\textbf{GPU Baseline System}} \\

\hline
\textbf{Processor}                                                   & \begin{tabular}[c]{@{}l@{}} 2x Intel Xeon Gold 5118 12-core processor @ 2.30GHz\end{tabular}  \\ \hline
\textbf{Main Memory}                                                        & \begin{tabular}[c]{@{}l@{}}\SI{512}{\giga\byte} total capacity\\16$\times{}$\SI{32}{\giga\byte} DDR4 (RDIMMs)\end{tabular}  \\ \hline
\textbf{GPU}                                                        & \begin{tabular}[c]{@{}l@{}} 1$\times{}$ NVIDIA A100 (PCIe, \SI{80}{\giga\byte})\end{tabular}  \\ \hline
\end{tabular}
}
\end{table}

\subsection{Baseline Implementations}
\label{sec:baseline_implementations}

\noindent\textbf{CPU Baseline \sr{Implementation}.}  \hluo{We implement our CPU baselines using PyTorch~\cite{paszke2019pytorch}. We implement three distributed optimization algorithms, MA-SGD, GA-SGD, and ADMM, to train LR and SVM models, using the optimizers and communication libraries provided by PyTorch~\cite{paszke2019pytorch}. We consider each CPU thread as a \emph{worker} in the distributed optimization algorithms.}

\noindent\textbf{GPU Baseline \sr{Implementation}.} \hluo{We implement our GPU baselines using PyTorch~\cite{paszke2019pytorch}.} We only implement mini-batch SGD on the GPU because PyTorch does not provide a way to limit the amount of GPU resources the kernels use, causing model averaging to be serialized on a single GPU. \hluo{\om{For fair comparison,} we do not use a cluster of GPUs for our baseline because the UPMEM \srm{PIM} system is a single-server node.\footnote{\srm{A multi-GPU system can be compared with a multi-UPMEM PIM system, which we leave for future work.}}}

\subsection{Experiment Implementation Details}
\label{sec:experimental_settings}

\noindent\textbf{Data Format.} Quantization is a popular approach used in the ML community to enable fixed-point operations~\cite{choquette2021nvidia,dettmers2022gpt3,xi2023training}. We conduct ML training on \srm{the \emph{real-world} UPMEM PIM system} on quantized~\cite{wang2019accelerating,chung2018serving,han2015deep,umuroglu2017finn} training data and models. \om{B}oth \om{data and models are} represented \om{using} a 32-bit fixed-point format because \om{the} UPMEM PIM system does not natively support floating-point operations. We use \om{the FP32} floating-point format for our CPU and GPU baselines because 1) CPUs and GPUs natively support it and 2) it provides higher accuracy.

\noindent\textbf{Hyperparameter Tuning.} We tune the learning rates and regularization terms for all workloads \om{we evaluate}. We open source all tested hyperparameters along with our complete codebase \srm{at https://github.com/CMU-SAFARI/PIM-Opt}.

\noindent\textbf{Regularization.} We use standard regularization techniques to achieve lower generalization error\srm{s}. For MA-SGD and GA-SGD, we add a\om{n} $L^2$ regularization term to the loss functions of the LR and SVM models. For ADMM, we include $L^2$ regularization for the SVM loss function, while we include $L^1$ regularization for LR. By including $L^1$ regularization for LR ADMM, the dual optimization problem admits a closed-form solution similar to SVM ADMM with $L^2$ regularization~\cite{MAL-016_Boyd_ADMM}.

\noindent\textbf{Batch Size.} \sr{Given a batch of size $b$, for both MA-SGD and ADMM, each \emph{worker} processes $b$ samples in each iteration of mini-batch SGD. In contrast, for GA-SGD running on a system consisting of $N$ \emph{workers}, each \emph{worker} processes $b/N$ samples before the intermediate gradients are communicated with the parameter server. For simplicity, assume that $b$ is divisible by the number of \emph{workers} $N$.} \om{When} training models on \om{the} YFCC100M-HNfc6 \om{dataset}, we consider batch sizes $8$,~$16$,~$32$, and $64$ for MA-SGD/ADMM and $4096$ ($4$K), 8192 ($8$K), 16'384 ($16$K), and 32'768 ($32$K) for GA-SGD. \om{When} training models on Criteo, we consider batch sizes 1024 ($1$K), 2048 ($2$K), 4096 ($4$K), and 8192 ($8$K) for MA-SGD/ADMM, and 131'072 ($131$K), 262'144 ($262$K), 524'288 ($524$K), and 1'048'576 ($1048$K) for GA-SGD. We use different batch sizes for Criteo due to its orders of magnitude larger number of samples in the training dataset (\secref{sec:datasets}). \srmf{For each experiment in~\secref{sec:evaluation} (except for the batch size sensitivity analysis), we tune the batch size to ensure high accuracy, high performance in terms of total training time, and fair comparison of algorithms \& architectures.}

\noindent\textbf{Initialization.} \hluo{For both the UPMEM PIM \sr{system} implementation and the CPU/GPU baselines, the training data and model weights \emph{initially} \om{reside} in main memory. For \sr{the} UPMEM PIM \sr{system} and GPU experiments, the initialization \om{phase} includes transferring the data from the main memory to \om{the PIM DRAM bank and the GPU global memory.}}

\subsection{Datasets}
\label{sec:datasets}
We consider two large-scale datasets, YFCC100M-HNfc6 and Criteo 1TB Click Logs (Criteo).

\noindent1) \textbf{YFCC100M-HNfc6}~\cite{amato2016yfcc100m} consists of 97M samples with features extracted by a deep convolutional neural network from the YFCC100M multimedia dataset~\cite{yfcc100m}. Each sample consists of $4096$ floating-point dense features and a collection of tags. We randomly sample and shuffle data points with the tag \emph{"outdoor"}, treating them as positive labels, and sample the same number of data points with the tag \emph{"indoor"}, treating them as negative labels, turning this subset into a binary classification task. Although SGD randomly draws samples in theory, in practice, it is common to randomly shuffle the training dataset and sequentially process training samples at runtime, which generally is much faster~\cite{bottou2012stochastic}. We apply standard normalization to each feature column, and for our implementation on the \srm{UPMEM PIM system}, we quantize the normalized dataset into a 32-bit fixed-point format. The total size of model parameters is $4$KB.

\noindent2) \textbf{Criteo 1TB Click Logs \sr{(Criteo)}}~\cite{criteo1tb2014} preprocessed by LIBSVM~\cite{LIBSVMData2023} \srmf{consists} of approximately $4.37$ billion high-dimensional sparse samples with $1$M features. Criteo is a popular click-through rate prediction dataset. Data points labeled \emph{"click"} are treated as positive and \emph{"no-click"} as negative labels. The dataset has been collected over $24$ days and is highly imbalanced, with only $0.034$\% of data points being \emph{"clicks"}. To construct the training dataset, we randomly sample and shuffle from day $0$ to $22$ while maintaining the ordering only among days. We \om{use} the entire day $23$ as a test dataset for all our experiments. Each data point consists of a label and $39$ categorical features representing a sparse embedding in a $1$M-dimensional feature space. \om{W}hile data points only consist of $40$ parameters, the models/gradients consist of $1$M variables and, therefore, incur a significantly higher communication overhead compared to YFCC100M-HNfc6. We use the \sr{area under the receiver operating characteristics curve (AUC score)~~\cite{huang2005using}} to assess the generalization capabilities of models trained on Criteo due to its imbalanced data distribution. The total size of the model parameters is $4$MB.

Table~\ref{tab:dataset} summarizes the dataset configurations used in our experiments for \sr{the} scaling analysis and comparison to the CPU \om{and GPU} baseline system\om{s}.

\begin{table}[h]
\centering
\caption{Dataset \sr{C}onfiguration\om{s}}
\label{tab:dataset}
\resizebox{\linewidth}{!}{

\begin{tabular}{ccccc}
\hline
\hline
\multicolumn{5}{c}{\textbf{YFCC100M-HNfc6}} \\
\hline
\# Workers                  & \# Training samples                               & Training size (GB) & \# Test samples                   & Test size (GB) \\ \hline
256 DPUs  & 851'968                                        & 13.96                       & 212'992                           & 3.49                    \\ 
512 DPUs  & 1'703'936          & 27.92    & 425'984 & 6.98   \\ 
1024 DPUs  & 3'407'872        & 55.83    & 851'968 & 13.96   \\ 
2'048 DPUs  & 6'815'744          &  111.67    &  1'703'936 &  27.92   \\
128 CPU threads & 6'815'744                                      & 111.67                      & 1'703'936                         & 27.92                   \\ 
1 GPU & 6'815'744                                      & 111.67                      & 1'703'936                         & 27.92                   \\
\hline
\hline
\multicolumn{5}{c}{\textbf{Criteo}} \\
\hline
\# Workers                  & \# Training samples                               & Training size (GB) & \# Test samples                   & Test size (GB) \\ \hline
256 DPUs  & 50'331'648                                     & 8.05                        & 178'236'537                       & 28.52                   \\ 
512 DPUs  & 100'663'296 & 16.11      & 178'236'537                       & 28.52                   \\ 
1'024 DPUs  & 201'326'592 & 32.21      & 178'236'537                       & 28.52                   \\ 
2'048 DPUs  & 402'653'184 & 64.42      & 178'236'537                       & 28.52                   \\ 
128 CPU threads & 402'653'184                                    & 64.42                       & 178'236'537                       & 28.52                   \\ 
1 GPU & 402'653'184                                    & 64.42                       & 178'236'537                       & 28.52                   \\ \hline
\end{tabular}
}
\end{table}
\section{Evaluation}
\label{sec:evaluation}

\noindent We evaluate ML training of 1) dense models on the YFCC100M-HNfc6 dataset (\secref{sec:experiments_YFCC100M}), and 2) high-dimensional sparse models on the Criteo 1TB Click Logs dataset (\secref{sec:experiments_Criteo}). \sraf{We do the following analyses.}

    \noindent \textbf{PIM Performance Breakdown.} To understand key characteristics of distributed \srmf{ML} on \om{the} UPMEM PIM system \srf{using 2048 DPUs}, we break the training time of one global epoch down into \sraf{1)} communication \srf{and} synchronization between PIM and the parameter server (\texttt{\srmf{Comm./Sync. Para. Server}}, \srmf{i.e., the fraction of training time for communicating gradients and models, and worker synchronization; see \secref{sec:implementation} for details), 2) PIM computation time (\texttt{\srmf{PIM Comp.}}, \srf{i.e., the fraction of training time to execute arithmetic operations by UPMEM PIM processing units)}, and 3) PIM data movement time \srf{(i.e., the fraction of training time for data movement between MRAM and WRAM)}}.

    \noindent \textbf{Algorithm Selection.} To show \srof{that it is} importan\srof{t} \srof{to} carefully \emph{choos\srof{e}} the distributed optimization algorithm that best fits \srmf{the UPMEM PIM system}, we compare the total training time and \srf{the test accuracy (AUC score\srmf{; see~\secref{sec:datasets}})} \srf{to perform ML training on the dataset YFCC100M-HNfc6 (Criteo)} for several combinations of models \srf{(i.e., LR, SVM)}, algorithms \srf{(i.e., MA-SGD, GA-SGD, ADMM, mini-batch SGD)}, and architecture\srf{s} \srf{(i.e., UPMEM PIM system, CPU \sraf{baseline system},\srmf{ and} GPU \sraf{baseline system})}.

    \noindent \textbf{Batch Size.} \sraf{We} study the impact \srmf{of the batch size} on performance in terms of total training time and \srf{the test accuracy (AUC score)} \srf{to perform ML training on the dataset YFCC100M-HNfc6 (Criteo)}. \sraf{We analyze several} batch sizes on the \srmf{UPMEM} PIM \srmf{system} and the CPU \sraf{baseline system}. \srmf{We only implement mini-batch SGD on the GPU because PyTorch does not provide a way to limit the amount of GPU resources the kernels use, causing model averaging to be serialized on a single GPU (see \secref{sec:baseline_implementations}).}

    \noindent \textbf{Scaling.} We explore two different scaling variants to assess the impact of scaling on total training time \srf{(i.e., \srmf{for} 10 global epochs)} and \srf{test} accuracy \srf{(AUC score)} \srmf{on the UPMEM PIM system} \srf{for the dataset YFCC100M-HNfc6 (Criteo)}. 1) \textit{Weak Scaling.} We increase the number of DPUs \sraf{from $256$ to $2048$} in our experiments while the training dataset size is increased \sraf{from $13.96$GB to $111.67$GB for YFCC100M-HNfc6 \srmf{(}from $8.05$GB to $64.42$GB for Criteo\srmf{)}}. 2) \textit{Strong Scaling.} We fix the training dataset size that fits on the smallest number of DPUs (i.e., $256$). \sraf{As we scale the number of DPUs from $256$ to $2048$ the training dataset remains unchanged, i.e., $13.96$GB for YFCC100M-HNfc6 \srmf{(}$8.05$GB for Criteo\srmf{)}}.

\subsection{Evaluation of YFCC100M-HNfc6}
\label{sec:experiments_YFCC100M}

\begin{figure}[h]
    \centering
    \includegraphics[width=\linewidth]{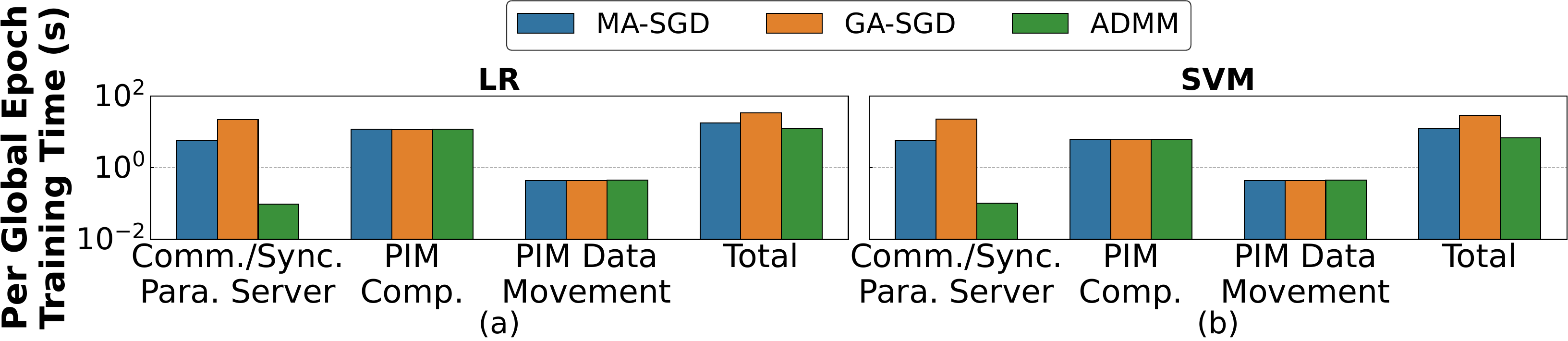}
    \caption{Per global epoch \sraf{training time} breakdown \srmf{into \texttt{\srmf{Comm./Sync. Para. Server}}, \texttt{\srmf{PIM Comp.}}, and PIM data movement time for LR (a) and SVM (b)}.}
    \label{fig:yfcc_perf_breakdown}
\end{figure}

\textbf{PIM Performance Breakdown.} In Fig.~\ref{fig:yfcc_perf_breakdown}, we \sraf{show} the training time for one global epoch (y-axis) and breakdown \sraf{the} training time into communication \srf{and} synchronization between PIM and the parameter server (\texttt{\srmf{Comm./Sync. Para. Server}}), PIM computation time (\texttt{\srmf{PIM Comp.}}), and PIM data movement time (x-axis) \sraf{for LR (\srmf{Fig.~\ref{fig:yfcc_perf_breakdown}(a)}) and SVM (\srmf{Fig.~\ref{fig:yfcc_perf_breakdown}(b)})}. For MA-SGD and ADMM, we set the batch size to $8$. For GA-SGD, we \sraf{set} the batch size to $4$K. 

\observation{\label{obsv:perf_breakdown_yfcc_1}Communication and synchronization between the parameter server and PIM is a bottleneck for MA-SGD/GA-SGD.}

For instance, LR MA-SGD (GA-SGD) communication and synchronization between PIM and the parameter server requires $56.0$x ($223.3$x) more time compared to ADMM. Here, we observe that \emph{communication-efficient} optimization algorithms such as ADMM improve performance.

\observation{\label{obsv:perf_breakdown_yfcc_2}For all combinations of optimization algorithms and models, PIM computation \sraf{takes more time than PIM data movement} on \srmf{the UPMEM} PIM.} 

For instance, LR (SVM) MA-SGD on PIM spends $26.75$x ($14.05$x) more time on computation than moving data between MRAM and WRAM. PIM spends less time on computation for SVM than LR because SVM \srf{exhibits} lower computational complexity.

\takeawaybox{\srmf{The UPMEM} PIM is \emph{less} suitable for ML models and optimization algorithms that require \srmf{frequent} communication \srf{and} synchronization \srf{between} PIM and the parameter server.}

\begin{figure}[h]
    \centering
    \includegraphics[width=\linewidth]{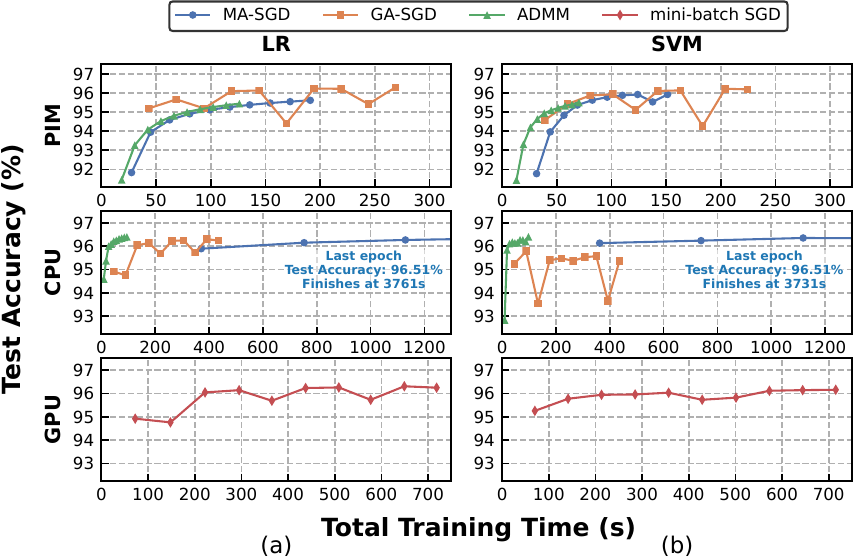}
    \caption{Comparison \srf{of} various models \srmf{(LR (a) and SVM (b))}, algorithms \srmf{(MA-SGD, GA-SGD, ADMM, and mini-batch SGD)}, and architectures \srmf{(PIM, CPU, and GPU)}. We study \om{the} test accuracy (\srf{\hluof{at} the last global epoch}) and total training time \srf{(10 global epochs).}}
    \label{fig:yfcc_perf_comparison}
    \vspace{-0.3em}
\end{figure}

\noindent\textbf{Algorithm Selection.} In Fig.~\ref{fig:yfcc_perf_comparison}, we study the test accuracy (\srf{i.e., reached in the last global epoch;} y-axis) and total training time (\srf{i.e., \srmf{for} 10 global epochs;} x-axis) \srf{with LR (\srmf{Fig.~\ref{fig:yfcc_perf_comparison}(a)}) and SVM (\srmf{Fig.~\ref{fig:yfcc_perf_comparison}(b)}}). \srmf{The UPMEM} PIM \srmf{system} with $2048$ DPUs (top), the CPU \sraf{baseline system} (middle), and the GPU \sraf{baseline system} (bottom). For the algorithms MA-SGD and ADMM, we set the batch size to $8$. For GA-SGD and \srf{mini-batch} SGD, \sraf{we set the batch size to} $4$K.

\observation{\label{obsv:yfcc_perf_comparison} \srmf{The difference in total training time between MA-SGD and ADMM is significantly lower on \om{the} \srmf{UPMEM} PIM compared to the CPU. GA-SGD is slower than ADMM for all configurations of LR, SVM, \srf{the} \srmf{UPMEM} PIM, and \srf{the} CPU.}}

For example, on \srmf{the UPMEM} PIM \srmf{system} (CPU \sraf{baseline system}), we observe a speedup of $1.51$x ($39.79$x) with LR ADMM compared to LR MA-SGD. The higher speedup on the CPU \srf{\sraf{baseline system}} \srmf{is due to} the smaller number of workers \srf{and, therefore\srf{,} less communication overhead} compared to \srmf{the UPMEM} PIM \srmf{system}. For instance, on \srmf{the UPMEM} PIM \srmf{system} (CPU \sraf{baseline system}), we observe speedups of $3.19$x ($4.45$x) with SVM ADMM compared to SVM GA-SGD. This is a result of efficient communication for ADMM since local models are collected only after each DPU/CPU thread has processed its complete partition of the training dataset\srmf{.} \srmf{In contrast, for} GA-SGD, gradients are communicated \srmf{in each iteration}. This \srmf{can cause a very large communication overhead, especially when training large-scale models.}

\observation{\label{obsv:yfcc_ga_sgd_pim_vs_cpu}GA-SGD on \srmf{the UPMEM} PIM outperforms GA-SGD on the CPU and \srf{mini-batch} SGD on the GPU for both LR and SVM.}

For LR (SVM), GA-SGD on \srmf{the UPMEM} PIM \srmf{system} achieves speedups of $1.62$x ($1.94$x) over the CPU \sraf{baseline system} and $2.67$x ($3.19$x) over the GPU \srf{\sraf{baseline system}} running \srf{mini-batch} SGD. A possible explanation for these speedups is that per CPU thread, there is not enough work before synchronization with the parameter server (Obsv.~\ref{obsv:yfcc_perf_comparison}), and the batch size is too small on the GPU \srf{\sraf{baseline system}}. The \srmf{difference in the increase in training time} between LR and SVM results from SVM's lower computational complexity compared to LR, and therefore, in general, SVM is faster than LR on \srmf{the UPMEM} PIM \srmf{system}. 

\observation{\label{obsv:yfcc_admm_pim_vs_cpu}ADMM is faster on \om{the} \srmf{UPMEM} PIM for SVM compared to the CPU. For LR, the CPU is faster.}

For SVM ADMM, we observe a speedup of $1.39$x on \srmf{the UPMEM} PIM \srmf{system} compared to the CPU \sraf{baseline system}. In contrast, for LR ADMM, we notice a slowdown by a factor of $1.33$x on \srmf{the UPMEM} PIM \srmf{system} compared to the CPU \sraf{baseline system}. This is \srmf{expected} since the training of SVM on \srmf{the UPMEM} PIM \srmf{system} requires less computation and no lookup to approximate the sigmoid function compared to LR.

\takeawaybox{\srmf{The UPMEM} PIM is a viable alternative to the CPU and the GPU for training small dense models on large-scale datasets.}

\begin{figure}[h]
    \centering
    \includegraphics[width=\linewidth]{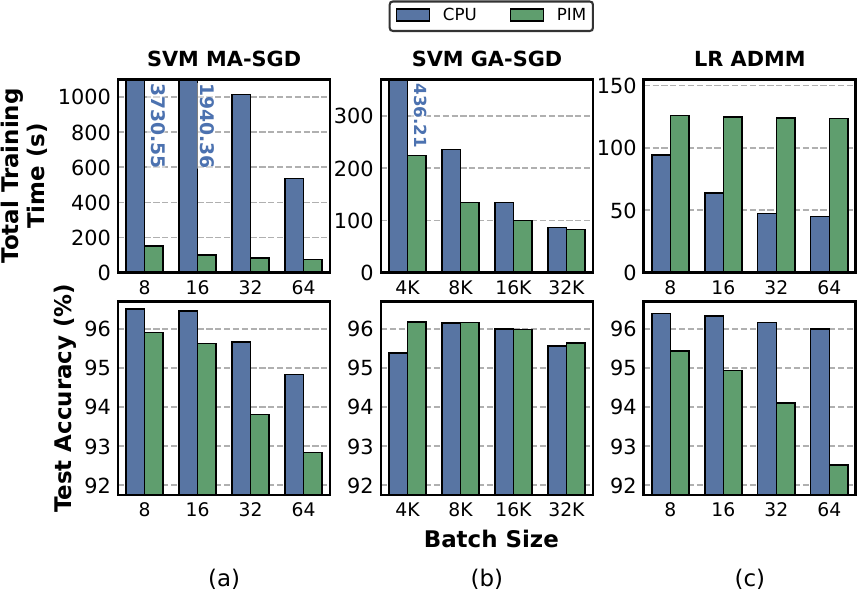}
    \caption{Impact \sraf{of batch size on} total training time \srf{(10 global epochs)} and test accuracy \srf{(\hluof{at} the last global epoch)} \srmf{for SVM MA-SGD (a), SVM GA-SGD (b), and LR ADMM (c)}.}
    \label{fig:yfcc_batch_size_sweep_compact}
\end{figure}

\noindent\textbf{Batch Size.} In Fig.~\ref{fig:yfcc_batch_size_sweep_compact}, we compare the total training time for 10 global epochs (\srf{y-axis;} first row), the test accuracy reached in the last global epoch (y-axis\srf{;} second row), and varying batch size (x-axis). \om{W}e illustrate a fixed combination of the model and optimization algorithm \srmf{for SVM MA-SGD (Fig.~\ref{fig:yfcc_batch_size_sweep_compact}(a)), SVM GA-SGD (Fig.~\ref{fig:yfcc_batch_size_sweep_compact}(b)), and LR ADMM (Fig.~\ref{fig:yfcc_batch_size_sweep_compact}(c))}. Each subplot compares \srmf{the UPMEM} PIM \srmf{system} with $2048$ DPUs and the CPU \sraf{baseline system} with $128$ CPU threads for every batch size.

\observation{\label{obsv:yfcc_batch_size_1}\srmf{As batch size increases, for MA-SGD and ADMM, we observe a reduction in the total training time on the CPU. In contrast, on the UPMEM PIM, the reduction of total training time is less significant.}}

When batch size \srmf{increases from $8$ to $64$}, the total training time of SVM MA-SGD on \srmf{the UPMEM} PIM \srmf{system} \srmf{decreases} by $2.01$x \srmf{from $151.36$s to $75.28$s}, compared to the CPU \srf{\sraf{baseline system}}, where the total training time \srmf{decreases} by $6.96$x \srmf{from $3730.55$s to $536.30$s}. \srf{This speedup is attributed to the fact that} larger batch sizes directly result in less communication. The reason for the high speedup on the CPU \srf{\sraf{baseline system}} is discussed in Obsv.~\ref{obsv:yfcc_perf_comparison}. For LR ADMM, the total training time \srmf{decreases} by $1.02$x on \srmf{the UPMEM} PIM \srmf{system}, compared to $2.10$x on the CPU \srf{\sraf{baseline system}}, respectively. \srf{This stems from the fact that the} local model update on PIM is not significantly more compute-intensive and only requires reading the gradient into WRAM compared to a single gradient step. On the CPU \srf{\sraf{baseline system}}, the slowdown likely stems from polluted caches due to the local model and the gradient to be loaded into the cache with an increased frequency of model updates for smaller batch sizes. We observe that \srmf{the} \srf{test} accuracy \srmf{decreases} as batch size \srmf{increases from $8$ to $64$}, e.g., for SVM MA-SGD, the test accuracy \srmf{decreases} from $95.92$\% to $92.84$\% on \srmf{the UPMEM} PIM \srmf{system} and from $96.51$\% to $94.83$\% on the CPU \srf{\sraf{baseline system}}. This \srmf{decrease} arises from the famous bias-variance tradeoff~\cite{geman1992neural} when training ML models, i.e., we want to reduce variance by increasing the batch size until we observe a drop in accuracy. Note that SGD-based algorithms admit unbiased gradient estimates~\cite{goodfellow2016}. The discrepancy in test accuracy between \srmf{the UPMEM} PIM \srmf{system} and CPU \srf{\sraf{baseline system}} stems from the quantization of the training data and the model and a \srf{significantly} larger number of models on \srmf{the UPMEM} PIM \srmf{system}.

\observation{\label{obsv:yfcc_batch_size_2}Both \srmf{the UPMEM} PIM and the CPU benefit from larger batch sizes for GA-SGD.}

Only for GA-SGD, for both \srmf{the UPMEM} PIM \srmf{system} and the CPU \sraf{baseline system}, we observe a significant reduction in total training time as we increase the batch size, while jointly, the test accuracy only slightly degrades. This behavior is explained by the reduction of communication for larger batch sizes since each DPU/CPU thread can process more samples before gradients need to be collected and the model is updated. GA-SGD’s test accuracy is less sensitive to increasing batch size because GA-SGD has only one model \srf{(see Obsv.~\ref{obsv:first_yfcc_strong_scaling} for more details)}.

\takeawaybox{\srmf{The UPMEM} PIM has benefits for 1) models that require smaller batch sizes \sraf{to achieve high accuracy}, and 2) algorithms that minimize inter-DPU communication \srf{via the parameter server}.}

\begin{figure}[h]
    \centering
    \includegraphics[width=\linewidth]{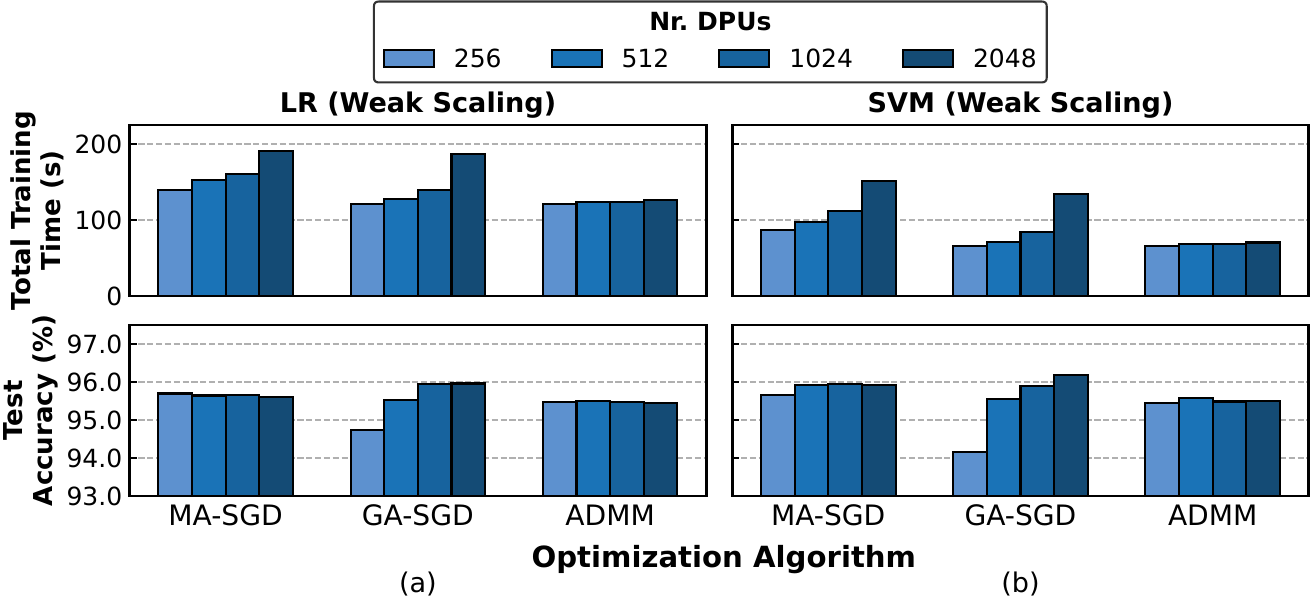}
    \caption{Impact of weak scaling on total training time \srf{(10 global epochs)} and test accuracy (\hluof{at} the last global epoch) \srmf{for LR (a) and SVM (b)}.}
    \label{fig:yfcc_weak_scaling}
\end{figure}

\noindent\textbf{Scaling.} In Fig.~\ref{fig:yfcc_weak_scaling}, \srmf{we study the weak scalability (i.e., the training dataset size increases proportionally as the number of DPUs increases) of using MA-SGD, GA-SGD, and ADMM (x-axis) to train LR (Fig.~\ref{fig:yfcc_weak_scaling}(a)) and SVM (Fig.~\ref{fig:yfcc_weak_scaling}(b)) models on the UPMEM PIM system. We plot the total training time for $10$ global epochs (\srf{y-axis;} first row), and the test accuracy reached in the last global epoch (y-axis\srf{;} second row). For all combinations of the models and optimization algorithms, we increase the number of DPUs from $256$ to $2048$ and proportionally increase the total training dataset size from from $13.96$GB to $111.67$GB.} For MA-SGD and ADMM, we \sraf{set} the batch size to $8$. For GA-SGD, we set the batch size to $8$K.

\observation{\label{obsv:yfcc_weak_scaling_1}\srmf{The UPMEM} PIM has good weak scalability with ADMM but \srf{poor} weak scalability with MA-SGD and GA-SGD in terms of total training time.}

As an example, for SVM ADMM (MA-SGD), we observe an increase of total training time by $1.08$x ($1.75$x), while the achieved test accuracy only changes very slightly as we scale from $256$ to $2048$ DPUs. The \srmf{increase in training time} is attributed to the slightly higher communication overhead for small, dense models as we scale the number of workers.

\observation{\label{obsv:yfcc_weak_scaling_2}\srmf{Among the algorithms we test, only GA-SGD's test accuracy consistently increases when both the training dataset size and the number of DPUs increase.}} 

For SVM GA-SGD, we observe an increase in total training time by $2.05$x, while the achieved test accuracy increases from $94.15$\% to $96.17$\% as we scale the number of DPUs from $256$ to $2048$. The slowdown stems from higher communication overhead when training with more DPUs. For GA-SGD, when we increase the number of \sraf{DPUs}, each DPU processes fewer samples per batch, exacerbating the communication overhead.

\begin{figure}[h]
    \centering
    \includegraphics[width=\linewidth]{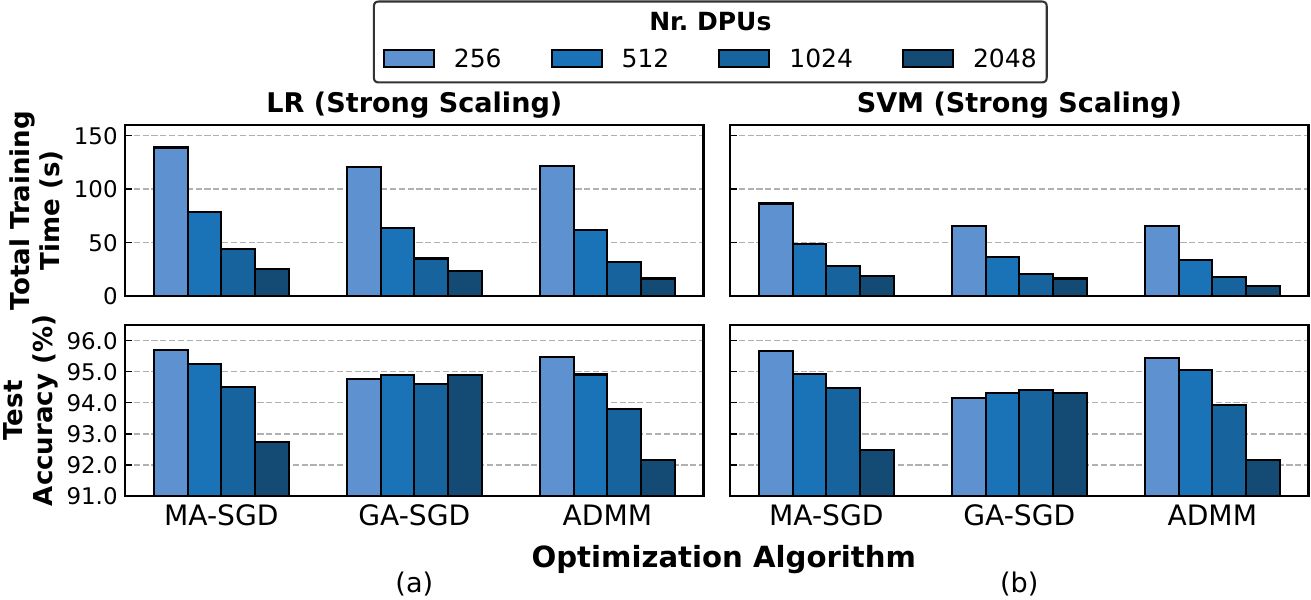}
    \caption{Impact of strong scaling on total training time \srf{(10 global epochs)} and test accuracy (\hluof{at} the last global epoch) \srmf{for LR (a) and SVM (b)}.}
    \label{fig:yfcc_strong_scaling}
\end{figure}

In Fig.~\ref{fig:yfcc_strong_scaling}, we use the same experiment setting as in Fig.~\ref{fig:yfcc_weak_scaling}, except that we fix the training dataset size as we scale the number of DPUs (strong scaling).

\observation{\label{obsv:first_yfcc_strong_scaling}\srmf{The UPMEM} PIM has good strong scalability in terms of total training time, but \srf{poor} in accuracy.}

As an example, for LR ADMM (MA-SGD), we observe a speedup of $7.43$x ($5.47$x), while the achieved test accuracy decreases from $95.46$\% ($95.70$\%) to $92.17$\% ($92.73$\%), as we scale from $256$ to $2048$ DPUs. In contrast, for LR GA-SGD, we observe a speedup of total training time by $5.22$x, while the achieved test accuracy only slightly improves as we scale the number of DPUs from $256$ to $2048$. Therefore, the \emph{communication-efficient} ADMM algorithm achieves a higher speedup compared to MA-SGD and GA-SGD. The observed reduction in test accuracy for a larger number of DPUs directly corresponding to a larger number of models when training ML models with MA-SGD and ADMM \srof{is due to the fact that more workers increase staleness as each worker uses its own local model before synchronizing with the parameter server.}\footnote{\om{For a theoretical analysis of this phenomenon of how the number of workers affects the convergence rate, we refer the reader to~\cite{zhou2017convergence}.}} Other works also make this empirical observation that convergence becomes slower as the number of workers is scaled~\cite{zhang2019mllib,wortsman2022fi}. However, these \sraf{works} consider a substantially smaller number of workers \srf{(i.e., up to $128$)}.

\takeawaybox{\label{take:4}\srf{The} scalability potential \srmf{of \om{the} UPMEM PIM} for training \srf{small dense models} is limited by its lack of \srf{direct} inter-DPU communication.}

\subsection{Evaluation of Criteo}
\label{sec:experiments_Criteo}

\begin{figure}[h]
    \centering
    \includegraphics[width=\linewidth]{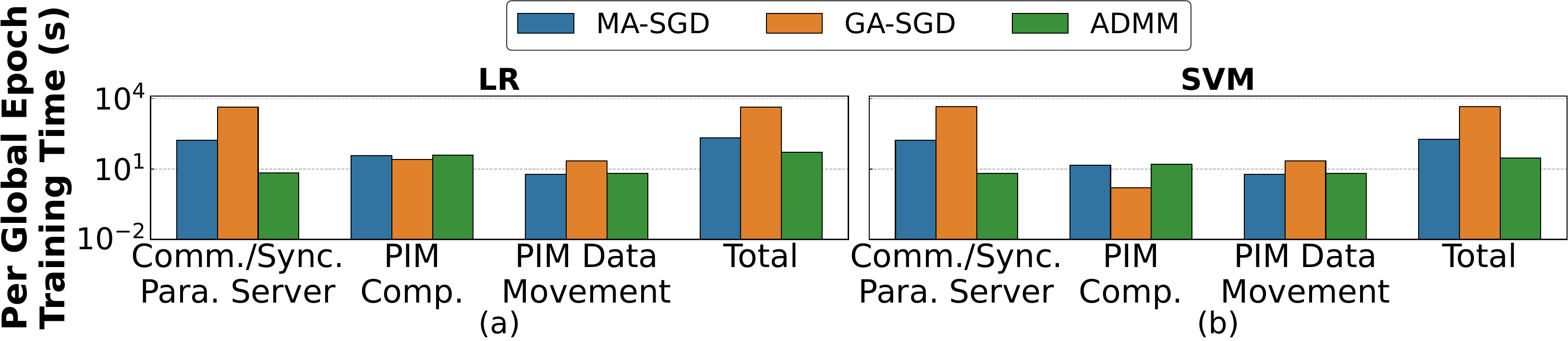}
    \caption{Per global epoch \sraf{training time} breakdown \srmf{into \texttt{\srmf{Comm./Sync. Para. Server}}, \texttt{\srmf{PIM Comp.}}, and PIM data movement time for LR (a) and SVM (b)}.}
    \vspace{0.2em}
    \label{fig:criteo_perf_breakdown}
\end{figure}

\textbf{PIM Performance Breakdown.} In Fig.~\ref{fig:criteo_perf_breakdown}, we \sraf{show} the training time for one global epoch (y-axis) and breakdown \sraf{the} training time into communication and synchronization between PIM and the parameter server (\texttt{\srmf{Comm./Sync. Para. Server}}), PIM computation time (\texttt{\srmf{PIM Comp.}}), and PIM data movement time (x-axis) \sraf{for LR (\srmf{Fig.~\ref{fig:criteo_perf_breakdown}(a)}) and SVM (\srmf{Fig.~\ref{fig:criteo_perf_breakdown}(b)})}. For MA-SGD and ADMM, we set the batch size to $2$K. For GA-SGD, we \sraf{set} the batch size to $262$K.

\observation{Communication and synchronization between PIM and the parameter server is a bottleneck for MA-SGD/GA-SGD.}

For instance, LR MA-SGD (GA-SGD) communication and synchronization between PIM and the parameter server requires $25.10$x ($640.35$x) more time compared to ADMM. This coincides with Obsv.~\ref{obsv:perf_breakdown_yfcc_1} for YFCC100M-HNfc6.

\observation{For \srmf{both} MA-SGD and ADMM, PIM computation \sraf{takes more time than PIM data movement} on \srmf{the UPMEM} PIM. For GA-SGD, PIM data movement \sraf{takes more time than PIM computation} on \srmf{the UPMEM} PIM.}

As an example, LR (SVM) MA-SGD on PIM spends $6.38$x ($2.44$x) more time on computation than moving data between MRAM and WRAM. \srmf{Compared to LR, SVM's lower computational complexity causes it to spend less time doing computation.} In contrast to our Obsv.~\ref{obsv:perf_breakdown_yfcc_2} for YFCC100M-HNfc6, for Criteo, \srmf{SVM GA-SGD  spends $14.29$x more time on moving data between MRAM and WRAM compared to computation on the PIM system.} This is \srmf{because the gradient update of GA-SGD requires sequentially reading the complete gradient into the WRAM} and subsequently back to MRAM. For Criteo, we can take advantage of larger individual data transfers that are more efficient compared to YFCC100M-HNfc6. \srmf{Therefore, } most of the computation of updating the model is offloaded to the parameter server.

\takeawaybox{\srmf{The \srmf{UPMEM}} PIM is \srmf{\emph{less}} suitable for training high-dimensional sparse models and optimization algorithms that require 
 \srmf{frequent} communication and synchronization between PIM and the parameter server.}

\begin{figure}[h]
    \centering
    \includegraphics[width=\linewidth]{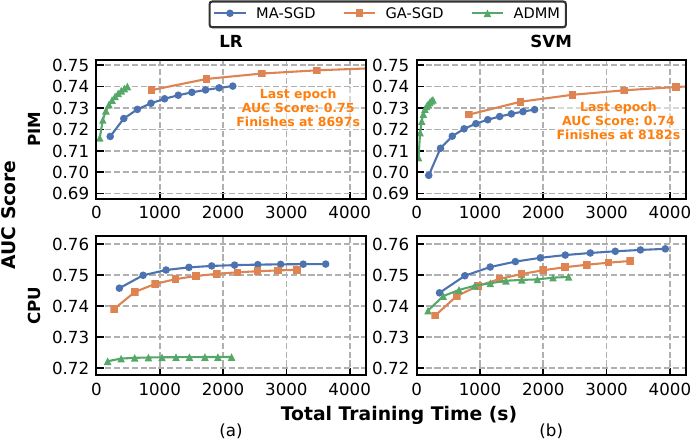}
    \caption{Comparison \srf{of} various models \srmf{(LR (a) and SVM (b))}, algorithms \srmf{(MA-SGD, GA-SGD, and ADMM)}, and architectures \srmf{(PIM and CPU)}. We study \om{the} AUC score (\srf{\hluof{at} the last global epoch}) and total training time \srf{(10 global epochs).}}
    \label{fig:criteo_perf_comparison}
\end{figure}

\noindent\textbf{Algorithm Selection.} In Fig.~\ref{fig:criteo_perf_comparison}, we study the AUC score (\srf{i.e., reached in the last global epoch;} y-axis) and total training time (\srf{i.e., \srmf{for} 10 global epochs;} x-axis) \srf{with LR (\srmf{Fig.~\ref{fig:criteo_perf_comparison}(a)}) and SVM (\srmf{Fig.~\ref{fig:criteo_perf_comparison}(b)}) \srmf{models}}. \srmf{The UPMEM} PIM \srmf{system} \srmf{uses} $2048$ DPUs (first row), \srmf{and} the CPU \sraf{baseline system} \srmf{uses} $128$ CPU threads (second row). For the algorithms MA-SGD and ADMM\srf{,} we set \srf{the} batch size to $2$K. For GA-SGD and \srf{mini-batch} SGD, \sraf{we set} the batch size to $524$K. For the GPU \sraf{baseline system}, we only report a per batch speedup comparison to PIM with GA-SGD (Obsv.~\ref{obsv:criteo_batch_gpu}) because \srf{the} training of Criteo's high-dimensional sparse model with \srf{mini-batch} SGD is \srmf{prohibitively} on the GPU \sraf{baseline system}.

\observation{ADMM outperforms MA-SGD for both LR and SVM on \om{the} \srmf{UPMEM} PIM \srmf{in both total training time and AUC score}. \srmf{On CPU}, ADMM outperforms MA-SGD in terms of total training time \srmf{but reaches a} lower AUC score.}

Training on the sparse dataset Criteo, for LR (SVM) ADMM, we observe a speedup of $4.40$x ($7.24$x) on \srmf{the UPMEM} PIM \srmf{system} and $1.70$x ($1.64$x) on the CPU \sraf{baseline system} compared to MA-SGD. \srmf{The reason for ADMM having a larger speedup compared to MA-SGD on the \srmf{UPMEM} PIM \srmf{system} than the CPU baseline is that the communication overhead of ADMM is much smaller compared to MA-SGD, which benefits \srmf{the UPMEM} PIM \srmf{system} more than the CPU.} \srmf{The reason for SVM to have a larger speedup than LR on the UPMEM PIM system is that SVM has a lower computational complexity compared to LR.}

\observation{ADMM and MA-SGD significantly outperform GA-SGD for both LR and SVM on \srmf{the {UPMEM}} PIM \srmf{in total training time with a negligible reduction in AUC score}. \srmf{On CPU}, ADMM outperforms GA-SGD only in terms of total training time. In contrast to \srmf{the UPMEM} PIM \srmf{system}, \srmf{on CPU}, MA-SGD is slightly slower than GA-SGD but achieves a higher AUC score.}

For instance, on \srmf{the UPMEM} PIM \srmf{system} (CPU \sraf{baseline system}), we observe speedups of $31.82$x ($1.41$x) with SVM ADMM compared to SVM GA-SGD at the cost of a reduction of the AUC score by $1.014$x ($1.007$x). The \srmf{difference in the increase in training time} between the \srmf{UPMEM} PIM \srmf{system} and CPU \sraf{baseline system} for GA-SGD is exacerbated \srmf{because there are more workers on the PIM system, causing} more intermediate gradients \srmf {to be} communicated over the slow channel between PIM and the parameter server. For SVM MA-SGD, we observe speedups of $4.39$x at the cost of a reduction of the AUC score by $1.02$x on \srmf{the \srmf{UPMEM}} PIM \srmf{system} compared to GA-SGD. In contrast, on the CPU \sraf{baseline system}, for SVM MA-SGD\srf{,} we observe a slowdown of $1.16$x and an increase of the AUC score by $1.01$x. The increase in training time on the CPU \sraf{baseline system} for MA-SGD compared to GA-SGD is a result of that for MA-SGD, each CPU thread needs to read its gradient into cache and update the model followed by communication of the models, while for GA-SGD, the intermediate gradients are communicated directly, and only a single model is updated. \srf{E}ach CPU thread processes the same number of samples for MA-SGD and GA-SGD.

\observation{\label{obsv:criteo_batch_gpu}GA-SGD on the CPU outperforms GA-SGD on \srmf{the UPMEM} PIM and the GPU.}

For LR (SVM), GA-SGD on the CPU \sraf{baseline system} achieves speedups of $2.75$x ($2.43$x) over the \srmf{UPMEM} PIM \srmf{system}. This observation differs from our Obsv.~\ref{obsv:yfcc_ga_sgd_pim_vs_cpu} for YFCC100M-HNfc6. The reason is that the communication overhead is exacerbated for Criteo because of the larger model size (i.e., $4$MB). For the GPU \sraf{baseline system} with \srf{mini-batch} SGD, we only report a per batch speedup comparison to \srmf{the UPMEM} PIM \srmf{system} with GA-SGD. Training of Criteo's high-dimensional sparse model is very slow on the GPU because only minimal computation is required for each sample. GA-SGD on \srmf{the UPMEM} PIM \srmf{system} achieves speedups of $10.65$x per batch for SVM over the GPU running \srf{mini-batch} SGD.

\observation{ADMM is faster on \srmf{the {UPMEM}} PIM for both LR and SVM compared to the CPU.}

For LR (SVM) ADMM, we observe speedups of $4.36$x ($9.33$x) on \srmf{the UPMEM} PIM \srmf{system} compared to the CPU \sraf{baseline system}. In contrast to the YFCC100M-HNfc6, we observe a speedup for LR when training on the Criteo dataset because\srf{,} per sample, there is less computation.

\takeawaybox{\srmf{The UPMEM PIM} is a viable alternative \srmf{to the CPU and the GPU} for training high-dimensional sparse models on large-scale datasets.}

\begin{figure}[h]
    \centering
    \includegraphics[width=\linewidth]{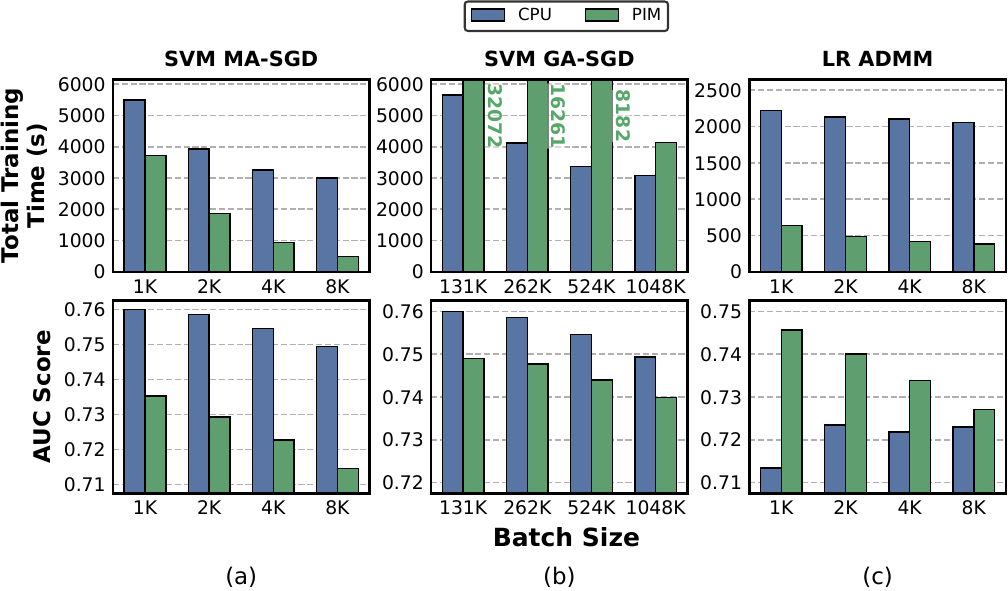}
    \caption{Impact \sraf{of batch size on} total training time \srf{(10 global epochs)} and AUC score \srf{(\hluof{at} the last global epoch)} \srmf{for SVM MA-SGD (a), SVM GA-SGD (b), and LR ADMM (c)}.}
    \vspace{-0.3em}
    \label{fig:criteo_batch_size_sweep_compact}
\end{figure}

\noindent\textbf{Batch Size.} In Fig.~\ref{fig:criteo_batch_size_sweep_compact}, we compare the total training time for 10 global epochs (\srf{y-axis;} first row), the AUC score reached in the last global epoch (y-axis\srf{;} second row), and varying batch size (x-axis). \om{W}e illustrate a fixed combination of the model and optimization algorithm \srmf{for SVM MA-SGD (Fig.~\ref{fig:criteo_batch_size_sweep_compact}(a)), SVM GA-SGD (Fig.~\ref{fig:criteo_batch_size_sweep_compact}(b)), and LR ADMM (Fig.~\ref{fig:criteo_batch_size_sweep_compact}(c))}. Each subplot compares \srmf{the UPMEM} PIM \srmf{system} with $2048$ DPUs and the CPU \sraf{baseline system} with $128$ CPU threads for every batch size.

\observation{As batch size \srmf{increases}, \srmf{both the UPMEM} PIM and the CPU exhibit performance \srmf{improvement} for MA-SGD. For ADMM, \srmf{increasing} the batch size only slightly \srmf{improves} performance for both \srmf{the UPMEM} PIM and the CPU.}

When batch size \srmf{increases from $1$K to $8$K}, the total training time of SVM MA-SGD on \srmf{the UPMEM} PIM \srmf{system} (CPU \sraf{baseline system}) decreases by $7.53$x ($1.83$x) \srmf{from $3725.71$s ($5488.46$s) to $494.98$s ($3004.8$s)}. \srmf{This is because larger batch sizes reduce the total amount of communication. The speedup on \srmf{the UPMEM} PIM \srmf{system} is higher compared to the CPU because the PIM system has more workers generating more communication, which benefits from the increase in batch size.}

\observation{Both \srmf{the UPMEM} PIM and the CPU benefit from larger batch sizes for training high-dimensional sparse models with GA-SGD.}

For \srmf{SVM} GA-SGD, for \srmf{the UPMEM} PIM \srmf{system} (CPU \sraf{baseline system}), we observe a reduction \srmf{by $7.53$x ($1.83$x)} in total training time as we increase the batch size while for both the AUC score only slightly degrades. This coincides with our Obsv.~\ref{obsv:yfcc_batch_size_2} for the YFCC100M-HNfc6 dataset.

\takeawaybox{When training high-dimensional sparse models, \srmf{the UPMEM} PIM has benefits for 1) models that are not sensitive to larger batch sizes, and 2) algorithms that require less inter-DPU communication \srf{via the parameter server}.}

\begin{figure}[h]
    \centering
    \includegraphics[width=\linewidth]{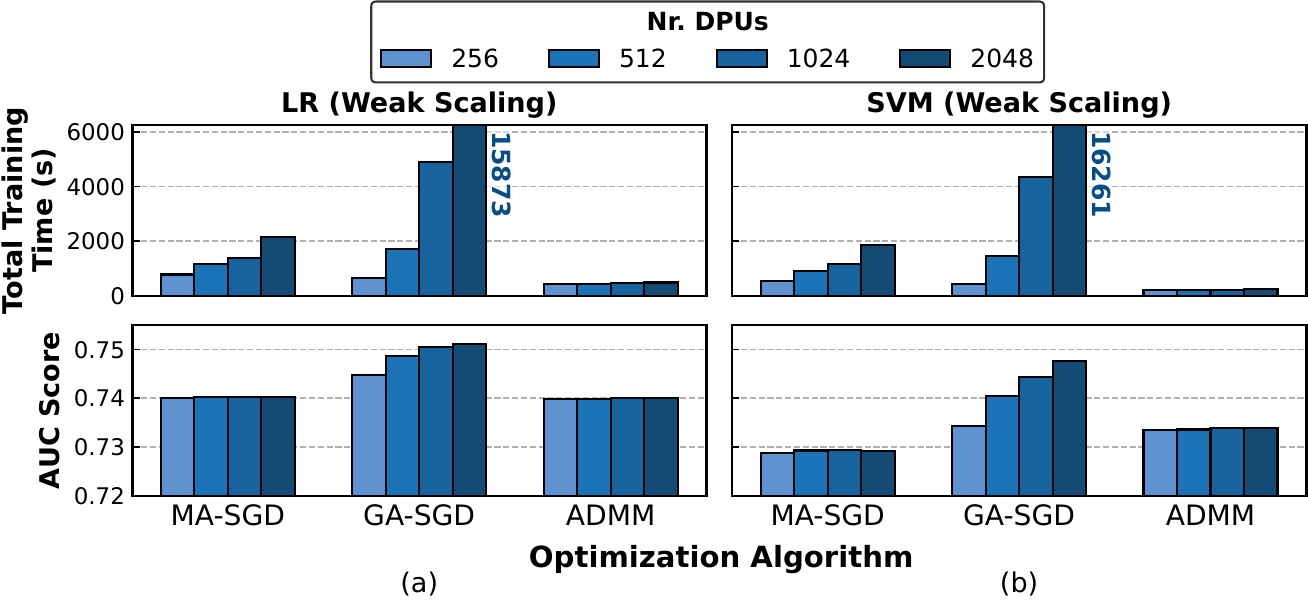}
    \caption{Impact of weak scaling on total training time \srf{(10 global epochs)} and AUC score (\hluof{at} the last global epoch) \srmf{for LR (a) and SVM (b)}.}
    \label{fig:criteo_weak_scaling}
\end{figure}

\noindent\textbf{Scaling.} In Fig.~\ref{fig:criteo_weak_scaling}, \srmf{we study the weak scalability (i.e., the training dataset size increases proportionally as the number of DPUs increases) of using MA-SGD, GA-SGD, and ADMM (x-axis) to train LR (\srmf{Fig.~\ref{fig:criteo_weak_scaling}(a)}) and SVM (\srmf{Fig.~\ref{fig:criteo_weak_scaling}(b)}) models on the UPMEM PIM system. We plot the total training time for $10$ global epochs (\srf{y-axis;} first row) and the AUC score reached in the last global epoch (y-axis\srf{;} second row). For all combinations of the models and optimization algorithms, we increase the number of DPUs from $256$ to $2048$ and proportionally increase the total training dataset size from $8.05$GB to $64.42$GB.} For MA-SGD and ADMM, we \srmf{set} the batch size to $2$K. For GA-SGD, we set the batch size to $262$K.

\observation{For high-dimensional sparse models, \srmf{UPMEM} PIM has good weak scalability with ADMM, but \srf{poor} weak scalability with MA-SGD and GA-SGD in terms of total training time.}

For example, for SVM ADMM (MA-SGD), we observe an increase of total training time by $1.28$x ($3.34$x), while the achieved AUC score changes very slightly as we scale from $256$ to $2048$ DPUs. \srmf{The difference in the increase in training time between ADMM and MA-SGD} is higher compared to YFCC100M-HNfc6 (i.e., see Obsv.~\ref{obsv:yfcc_weak_scaling_1}) because the communication overhead is exacerbated for Criteo's larger high-dimensional sparse model.

\observation{\srmf{Among the algorithms we test, only GA-SGD's AUC score consistently increases when both the training dataset size and the number of DPUs increase.}}

For SVM GA-SGD, we observe an increase of total training time by $37.82$x, while the achieved AUC score increases by $1.02$x as we scale the number of DPUs from $256$ to $2048$. The observations follow the same reasoning as in Obsv.~\ref{obsv:yfcc_weak_scaling_2} for the YFCC100M-HNfc6 dataset.

\begin{figure}[h]
    \centering
    \includegraphics[width=\linewidth]{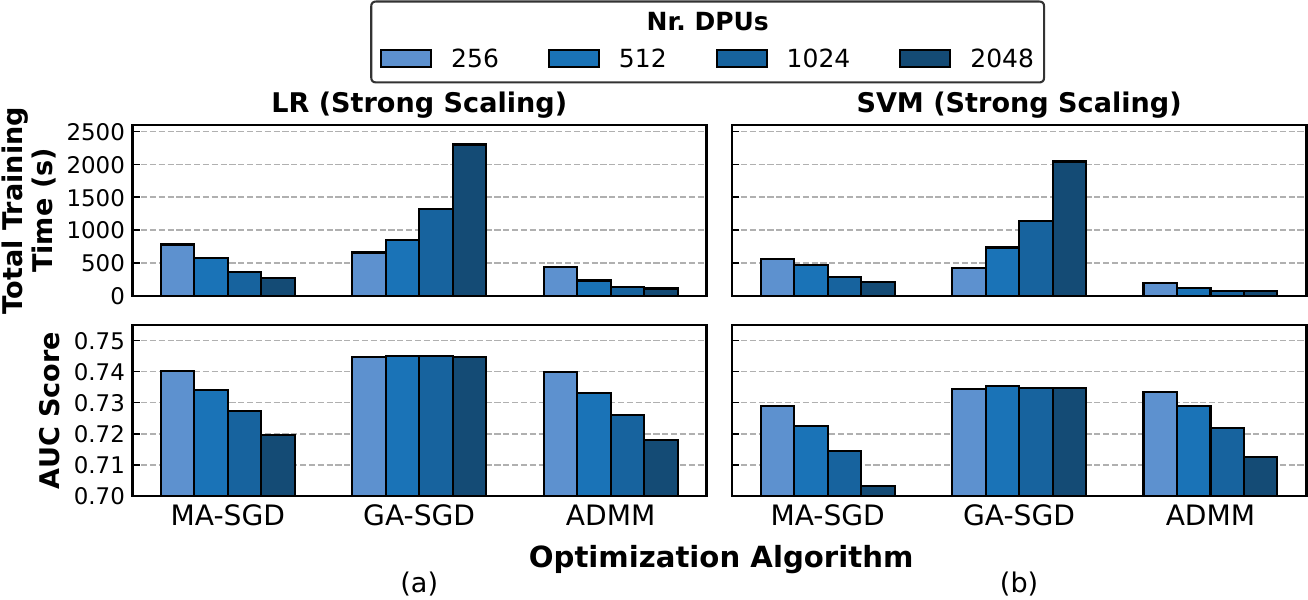}
    \caption{Impact of strong scaling on total training time \srf{(10 global epochs)} and AUC score (\hluof{at} the last global epoch) \srmf{for LR (a) and SVM (b)}.}
    \label{fig:criteo_strong_scaling}
\end{figure}

In Fig.~\ref{fig:criteo_strong_scaling}, we use the same experiment setting as in Fig.~\ref{fig:criteo_weak_scaling}, except that we fix the training dataset size as we scale the number of DPUs (strong scaling).

\observation{For high-dimensional sparse models, \srmf{the UPMEM} PIM has good strong scalability in terms of total training time, but \srf{poor} in terms of AUC score.}

For example, for LR ADMM (MA-SGD), we observe a speedup of $3.85$x ($2.87$x), while the achieved AUC score decreases from $0.74$ ($0.74$) to $0.718$ ($0.72$), as we scale from $256$ to $2048$ DPUs. In contrast, for LR GA-SGD, we observe an increase in total training time by $3.49$x, while the achieved AUC score changes only very slightly as we scale the number of DPUs from $256$ to $2048$. The \srmf{smaller} speedup of ADMM  and MA-SGD, and even a slowdown \om{for} GA-SGD compared to YFCC100M-HNfc6 (i.e., see Obsv.~\ref{obsv:first_yfcc_strong_scaling}), is a result of \srmf{the larger models in Criteo that induce more communication overhead between the DPUs and the parameter server.} The observed reduction of the AUC score follows the same line of reasoning as in the elaborations after Obsv.~\ref{obsv:first_yfcc_strong_scaling} \om{for} \srf{the} YFCC100M-HNfc6 dataset.

\takeawaybox{\label{take:8}\srf{The} scalability potential \srmf{of \om{the} UPMEM PIM} for training high-dimensional sparse models is limited by its lack of \srf{direct} inter-DPU communication.}
\section{Implications for \sr{PIM} Hardware Design}
\label{sec:implication}

\hluo{Our evaluation (\secref{sec:evaluation}) demonstrates that a \emph{real-world} PIM system (i.e., UPMEM) can be a viable alternative to state-of-the-art processor-centric architectures \sr{(e.g., CPU, GPU)} for memory-bound ML training workloads involving large-scale datasets if 1) the optimization algorithms are carefully chosen to fit \om{the} PIM architecture, and 2) the arithmetic operations and data types are natively supported by the PIM hardware.}

\hluo{From our observation\sr{s} and analyses, we argue the most fundamental architectural design changes that future PIM architectures (including both UPMEM and other devices that do not have on-chip interconnect between PIM processing units like Samsung HBM-PIM~\cite{kwon202125}, SK Hynix AiM~\cite{lee20221ynm}) should make to enable \om{fast and} efficient large-scale ML training is to enable more efficient communication among PIM processing units (e.g., DPUs in UPMEM) by adding interconnects and/or shared memory. \sr{Extending the UPMEM PIM system with on-chip interconnects (i.e., \om{direct} inter-DPU communication) enables the implementation of decentralized optimization algorithms \om{(e.g., \cite{aysal2009broadcast,boyd2005gossip,carli2010gossip})}. For instance, decentralized parallel SGD algorithms \om{\cite{lian2017can,lan2020communication,sirb2016consensus}} are a promising solution to overcome scalability challenges of the \emph{real-world} PIM system \om{because of their} two major advantages over its centralized counterpart\om{s}: 1) \om{decentralized} optimization algorithms significantly reduce communication on the busiest node, and 2) decentralized parallel SGD theoretically \om{provides} approximately linear speedup in terms of computational complexity as we scale the number of nodes~\cite{lian2017can}.} Without such \sr{on-chip interconnects}, the advantages of using decentralized optimization algorithms to accelerate large-scale ML training workloads~\cite{liu2020distributed,lian2017can,lan2020communication} \sr{cannot be leveraged and distributed ML workloads are} significantly limited because of the high synchronization and communication overhead\om{s} \sr{of centralized optimization algorithms}, as our Takeaways \sr{\ref{take:4} and \ref{take:8}} show.}

\hluo{For training workloads with even \om{larger} data and model sizes (e.g., training modern large language models with transformers on internet-scale text corpora~\cite{cotterell2023formal}), we propose to further reduce the synchronization and communication \om{overheads} of PIM architectures by enabling the separate allocation of the training data from the model into memory units. This is \sr{\om{because} the model is frequently accessed and updated during ML training. \om{In contrast,} the training data is \srm{less frequently} accessed as} it is common to train such models only for \srmf{a} few epochs~\cite{xue2024repeat} to reduce cost.}

We \om{posit} that a shift towards an algorithm-hardware codesign perspective is necessary in the context of ML training using PIM due to the high complexity of the design space, including algorithms, models, training, distributed system topology, and hardware design. With this \om{paper}, we hope to spark a \om{data-driven} discussion \om{and further research that can truly unleash} the full potential of \om{PIM on} ML training workloads.
\section{Related Work}

To our knowledge, this \om{paper} is the first to implement and rigorously evaluate distributed \om{\emph{Stochastic Gradient Descent} (SGD)} algorithms on a \emph{real-world} PIM system. \om{W}e describe other related works on \om{the} UPMEM PIM system, PIM for ML training and inference, and distributed \sr{optimization} algorithms.

\noindent\textbf{UPMEM PIM \sr{S}ystem.} \om{S}everal prior works \om{characterize} and overview the \om{UPMEM PIM} architecture \om{\cite{gomezluna2021prim,gomez2021benchmarking,nider2021case,devaux2019true,peccerillo2022survey,abumpimp2023}}. \om{Other} works explore a variety of algorithms and applications on \om{the} UPMEM PIM system, such as compiler\om{s} \sr{and} programming model\om{s}~\cite{chen2023simplepim,khan2022cinm}, libraries~\cite{10158230,giannoula2022sparsep}, simulation framework\om{s}~\cite{hyun2023pathfinding,hyun2024pathfinding}, bioinformatics~\cite{chen2023uppipe,diab2023framework,lavenier2020variant,lavenier2016blast}, security~\cite{gupta2023evaluating,jonatan2024scalability}, analytics \sr{and} databases~\cite{bernhardt2023pimdb,lim2023design,baumstark2023adaptive,baumstark2023accelerating,kang2023pim}, and ML training \sr{and} inference~\cite{gomez2022experimental,gomez2023evaluatingispass,giannoula2024accelerating,gogineni2024swiftrl,das2022implementation,Zarif_2023,kim2024optimal,wu2023pim}. \om{No} prior work examine\om{s} distributed \srm{SGD} algorithms commonly used for data-intensive ML training workloads on \om{the} UPMEM PIM system\om{.} 

\noindent\textbf{PIM for ML \sr{T}raining and \sr{I}nference.} There are several works on PIM acceleration for ML training~\cite{gao2015practical,falahati2018origami,vieira2018exploiting,sun2020one,shelor2019reconfigurable,saikia2019k}. However, none of these works \srm{use \emph{real-world} PIM systems}. Another body of \om{work}~\cite{deng2018dracc,boroumand2018google,gao2017tetris,boroumand2021google,cho2020mcdram,shin2018mcdram,azarkhish2017neurostream,kwon2019tensordimm,ke2020recnmp,cordeiro2021machine,lee2021task,park2021high,park2021trim,kim2020mvid,kwon202125,lee2021hardware,ke2021near,lee20221ynm,niu2022184qps} focuses on accelerating \om{ML} inference using PIM\srm{, showing the effectiveness of PIM at mitigating the data movement bottleneck in ML inference}. 

\noindent\textbf{Distributed \srm{Optimization} \sr{A}lgorithms.} \om{Various works}~\cite{wang2023cocktailsgd,xu2020compressed,alistarh2017qsgd,strom2015scalable} focus on algorithmically alleviating the communication overhead of centralized optimization algorithms since the parameter server has been identified to be the key bottleneck in distributed \sr{ML}. Other \om{works} \om{develop} decentralized optimization algorithms to minimize communication among many nodes~\cite{liu2020distributed,lian2017can,lan2020communication}. \srm{We believe that such algorithmic optimization combined with \om{enhanced} PIM hardware \om{(see \secref{sec:implication})} can fundamentally improve ML training performance on large-scale datasets.}
\section{Conclusion}

We evaluate and train ML models on large-scale datasets with centralized optimization algorithms on a \emph{real-world} PIM \srm{system} \sro{(i.e., UPMEM)}. We show \sro{that it is important to} carefully \emph{choose} the distributed optimization algorithm that fits \srm{the \emph{real-world}} PIM \srm{system} and analyze tradeoffs. We demonstrate that commercial general-purpose PIM systems can be a viable alternative \om{to processor-centric CPU and GPU architectures} for many ML training workloads on large-scale datasets. Our results demonstrate the necessity of adapting PIM architectures \om{to enable decentralized parallel SGD algorithms} to overcome scalability challenges for many distributed ML training workloads.

\begin{acks}
{
We thank the anonymous reviewers of PACT 2024 for feedback. We thank the SAFARI Research Group members for providing a stimulating intellectual environment. We thank UPMEM for providing hardware resources to perform this research. We acknowledge the generous gifts from our industrial partners, including Google, Huawei, Intel, and Microsoft. This work is supported in part by the Semiconductor Research Corporation (SRC), the ETH Future Computing Laboratory (EFCL), the European Union’s Horizon programme for research and innovation [101047160 - BioPIM], and the AI Chip Center for Emerging Smart Systems, sponsored by InnoHK funding, Hong Kong SAR (ACCESS).
}
\end{acks}

\balance
{
  \let\OLDthebibliography\thebibliography
  \renewcommand\thebibliography[1]{
    \OLDthebibliography{#1}
    \setlength{\parskip}{0pt}
    \setlength{\itemsep}{0pt}
  }
  \bibliographystyle{IEEEtranS}

\bibliography{refs}

\begin{thebibliography}{100}
\providecommand{\url}[1]{#1}
\csname url@samestyle\endcsname
\providecommand{\newblock}{\relax}
\providecommand{\bibinfo}[2]{#2}
\providecommand{\BIBentrySTDinterwordspacing}{\spaceskip=0pt\relax}
\providecommand{\BIBentryALTinterwordstretchfactor}{4}
\providecommand{\BIBentryALTinterwordspacing}{\spaceskip=\fontdimen2\font plus
\BIBentryALTinterwordstretchfactor\fontdimen3\font minus \fontdimen4\font\relax}
\providecommand{\BIBforeignlanguage}[2]{{%
\expandafter\ifx\csname l@#1\endcsname\relax
\typeout{** WARNING: IEEEtranS.bst: No hyphenation pattern has been}%
\typeout{** loaded for the language `#1'. Using the pattern for}%
\typeout{** the default language instead.}%
\else
\language=\csname l@#1\endcsname
\fi
#2}}
\providecommand{\BIBdecl}{\relax}
\BIBdecl

\bibitem{ahn2015scalable}
J.~Ahn, S.~Hong, S.~Yoo, O.~Mutlu, and K.~Choi, ``{A Scalable Processing-In-Memory Accelerator for Parallel Graph Processing},'' in \emph{{ISCA}}, 2015.

\bibitem{ahn2015pim}
J.~Ahn, S.~Yoo, O.~Mutlu, and K.~Choi, ``{PIM-Enabled Instructions: A Low-Overhead, Locality-Aware Processing-In-Memory Architecture},'' in \emph{{ISCA}}, 2015.

\bibitem{akin2015data}
B.~Akin, F.~Franchetti, and J.~C. Hoe, ``{Data Reorganization in Memory Using 3D-Stacked DRAM},'' in \emph{{ISCA}}, 2015.

\bibitem{alistarh2017qsgd}
D.~Alistarh, D.~Grubic, J.~Li, R.~Tomioka, and M.~Vojnovic, ``{QSGD: Communication-Efficient SGD via Gradient Quantization and Encoding},'' in \emph{{NeurIPS}}, 2017.

\bibitem{amato2016yfcc100m}
G.~Amato, F.~Falchi, C.~Gennaro, and F.~Rabitti, ``{YFCC100M-HNfc6: A Large-scale Deep Features Benchmark for Similarity Search},'' in \emph{{SISAP}}, 2016.

\bibitem{amdepyc2019}
{AMD}, ``{AMD EPYC 7742},'' \url{https://www.amd.com/en/support/downloads/drivers.html/processors/epyc/epyc-7002-series/amd-epyc-7742.html}, 2019.

\bibitem{asgari2021fafnir}
B.~Asgari, R.~Hadidi, J.~Cao, S.-K. Lim, and H.~Kim, ``{FAFNIR: Accelerating Sparse Gathering by Using Efficient Near-Memory Intelligent Reduction},'' in \emph{{HPCA}}, 2021.

\bibitem{asghari2016chameleon}
H.~Asghari-Moghaddam, Y.~H. Son, J.~H. Ahn, and N.~S. Kim, ``{Chameleon: Versatile and practical near-DRAM acceleration architecture for large memory systems},'' in \emph{{MICRO}}, 2016.

\bibitem{aysal2009broadcast}
T.~C. Aysal, M.~E. Yildiz, A.~D. Sarwate, and A.~Scaglione, ``{Broadcast Gossip Algorithms for Consensus},'' \emph{{IEEE Transactions on Signal processing}}, 2009.

\bibitem{azarkhish2017neurostream}
E.~Azarkhish, D.~Rossi, I.~Loi, and L.~Benini, ``{Neurostream: Scalable and Energy Efficient Deep Learning with Smart Memory Cubes},'' \emph{{TPDS}}, 2017.

\bibitem{baumstark2023accelerating}
A.~Baumstark, M.~A. Jibril, and K.-U. Sattler, ``{Accelerating Large Table Scan Using Processing-In-Memory Technology},'' \emph{{Datenbank-Spektrum}}, 2023.

\bibitem{baumstark2023adaptive}
------, ``{Adaptive Query Compilation with Processing-in-Memory},'' in \emph{{ICDEW}}, 2023.

\bibitem{bernhardt2023pimdb}
A.~Bernhardt, A.~Koch, and I.~Petrov, ``{pimDB: From Main-Memory DBMS to Processing-In-Memory DBMS-Engines on Intelligent Memories},'' in \emph{{DaMoN}}, 2023.

\bibitem{besta2021sisa}
M.~Besta, R.~Kanakagiri, G.~Kwasniewski, R.~Ausavarungnirun, J.~Ber{\'a}nek, K.~Kanellopoulos, K.~Janda, Z.~Vonarburg-Shmaria, L.~Gianinazzi, I.~Stefan \emph{et~al.}, ``{SISA: Set-Centric Instruction Set Architecture for Graph Mining on Processing-in-Memory Systems},'' in \emph{{MICRO}}, 2021.

\bibitem{boroumand2020practical}
A.~Boroumand, ``{Practical Mechanisms for Reducing Processor--Memory Data Movement in Modern Workloads},'' Ph.D. dissertation, {Carnegie Mellon University}, 2020.

\bibitem{boroumand2021google}
A.~Boroumand, S.~Ghose, B.~Akin, R.~Narayanaswami, G.~F. Oliveira, X.~Ma, E.~Shiu, and O.~Mutlu, ``{Google Neural Network Models for Edge Devices: Analyzing and Mitigating Machine Learning Inference Bottlenecks},'' in \emph{{PACT}}, 2021.

\bibitem{boroumand2018google}
A.~Boroumand, S.~Ghose, Y.~Kim, R.~Ausavarungnirun, E.~Shiu, R.~Thakur, D.~Kim, A.~Kuusela, A.~Knies, P.~Ranganathan \emph{et~al.}, ``{Google Workloads for Consumer Devices: Mitigating Data Movement Bottlenecks},'' in \emph{{ASPLOS}}, 2018.

\bibitem{boroumand2021polynesia}
A.~Boroumand, S.~Ghose, G.~F. Oliveira, and O.~Mutlu, ``{Polynesia: Enabling Effective Hybrid Transactional/Analytical Databases with Specialized Hardware/Software Co-Design},'' \emph{arXiv:2103.00798}, 2021.

\bibitem{boroumand2019conda}
A.~Boroumand, S.~Ghose, M.~Patel, H.~Hassan, B.~Lucia, R.~Ausavarungnirun, K.~Hsieh, N.~Hajinazar, K.~T. Malladi, H.~Zheng \emph{et~al.}, ``{CoNDA: Efficient Cache Coherence Support for Near-Data Accelerators},'' in \emph{{ISCA}}, 2019.

\bibitem{boroumand2016lazypim}
A.~Boroumand, S.~Ghose, M.~Patel, H.~Hassan, B.~Lucia, K.~Hsieh, K.~T. Malladi, H.~Zheng, and O.~Mutlu, ``{LazyPIM: An Efficient Cache Coherence Mechanism for Processing-in-Memory},'' \emph{{ICAL}}, 2016.

\bibitem{boser1992training}
B.~E. Boser, I.~M. Guyon, and V.~N. Vapnik, ``{A Training Algorithm for Optimal Margin Classifiers},'' in \emph{{COLT}}, 1992.

\bibitem{bottou2010large}
L.~Bottou, ``{Large-Scale Machine Learning with Stochastic Gradient Descent},'' in \emph{{COMPSTAT}}, 2010.

\bibitem{bottou2012stochastic}
------, \emph{{Stochastic Gradient Descent Tricks}}.\hskip 1em plus 0.5em minus 0.4em\relax Springer, 2012.

\bibitem{boyd2005gossip}
S.~Boyd, A.~Ghosh, B.~Prabhakar, and D.~Shah, ``{Gossip Algorithms: Design, Analysis and Applications},'' in \emph{{INFOCOM}}, 2005.

\bibitem{MAL-016_Boyd_ADMM}
S.~Boyd, N.~Parikh, E.~Chu, B.~Peleato, J.~Eckstein \emph{et~al.}, ``{Distributed Optimization and Statistical Learning via the Alternating Direction Method of Multipliers},'' \emph{{Foundations and Trends{\textregistered} in Machine Learning}}, 2011.

\bibitem{boyd2004convex}
S.~P. Boyd and L.~Vandenberghe, \emph{{Convex Optimization}}.\hskip 1em plus 0.5em minus 0.4em\relax {Cambridge University Press}, 2004.

\bibitem{cali2020genasm}
D.~S. Cali, G.~S. Kalsi, Z.~Bing{\"o}l, C.~Firtina, L.~Subramanian, J.~S. Kim, R.~Ausavarungnirun, M.~Alser, J.~Gomez-Luna, A.~Boroumand \emph{et~al.}, ``{GenASM: A High-Performance, Low-Power Approximate String Matching Acceleration Framework for Genome Sequence Analysis},'' in \emph{{MICRO}}, 2020.

\bibitem{carli2010gossip}
R.~Carli, F.~Fagnani, P.~Frasca, and S.~Zampieri, ``{Gossip Consensus Algorithms via Quantized Communication},'' \emph{{Automatica}}, 2010.

\bibitem{chang2017understanding}
K.~K. Chang, ``{Understanding and Improving the Latency of DRAM-based Memory Systems},'' Ph.D. dissertation, {Carnegie Mellon University}, 2017.

\bibitem{chang2016low}
K.~K. Chang, P.~J. Nair, D.~Lee, S.~Ghose, M.~K. Qureshi, and O.~Mutlu, ``{Low-Cost Inter-Linked Subarrays (LISA): Enabling fast inter-subarray data movement in DRAM},'' in \emph{{HPCA}}, 2016.

\bibitem{chen2023simplepim}
J.~Chen, J.~G{\'o}mez-Luna, I.~El~Hajj, Y.~Guo, and O.~Mutlu, ``{SimplePIM: A Software Framework for Productive and Efficient Processing-in-Memory},'' in \emph{{PACT}}, 2023.

\bibitem{chen2023uppipe}
L.-C. Chen, C.-C. Ho, and Y.-H. Chang, ``{UpPipe: A Novel Pipeline Management on In-Memory Processors for RNA-seq Quantification},'' in \emph{{DAC}}, 2023.

\bibitem{chi2016prime}
P.~Chi, S.~Li, C.~Xu, T.~Zhang, J.~Zhao, Y.~Liu, Y.~Wang, and Y.~Xie, ``{PRIME: A Novel Processing-In-Memory Architecture for Neural Network Computation in ReRAM-based Main Memory},'' in \emph{{ISCA}}, 2016.

\bibitem{chin2015fast}
W.-S. Chin, Y.~Zhuang, Y.-C. Juan, and C.-J. Lin, ``{A Fast Parallel Stochastic Gradient Method for Matrix Factorization in Shared Memory Systems},'' \emph{{TIST}}, 2015.

\bibitem{chin2015learning}
------, ``{A Learning-Rate Schedule for Stochastic Gradient Methods to Matrix Factorization},'' in \emph{{PAKDD}}, 2015.

\bibitem{cho2020mcdram}
S.~Cho, H.~Choi, E.~Park, H.~Shin, and S.~Yoo, ``{McDRAM v2: In-Dynamic Random Access Memory Systolic Array Accelerator to Address the Large Model Problem in Deep Neural Networks on the Edge},'' \emph{{IEEE Access}}, 2020.

\bibitem{choquette2021nvidia}
J.~Choquette, W.~Gandhi, O.~Giroux, N.~Stam, and R.~Krashinsky, ``{NVIDIA A100 Tensor Core GPU: Performance and Innovation},'' in \emph{{MICRO}}, 2021.

\bibitem{chung2018serving}
E.~Chung, J.~Fowers, K.~Ovtcharov, M.~Papamichael, A.~Caulfield, T.~Massengill, M.~Liu, D.~Lo, S.~Alkalay, M.~Haselman \emph{et~al.}, ``{Serving DNNs in Real Time at Datacenter Scale with Project Brainwave},'' in \emph{{MICRO}}, 2018.

\bibitem{cordeiro2021machine}
A.~S. Cordeiro, S.~R. dos Santos, F.~B. Moreira, P.~C. Santos, L.~Carro, and M.~A. Alves, ``{Machine Learning Migration for Efficient Near-Data Processing},'' in \emph{{PDP}}, 2021.

\bibitem{cortes1995support}
C.~Cortes, ``{Support-Vector Networks},'' \emph{{Machine Learning}}, 1995.

\bibitem{cotterell2023formal}
R.~Cotterell, A.~Svete, C.~Meister, T.~Liu, and L.~Du, ``{Formal Aspects of Language Modeling},'' \emph{arXiv:2311.04329}, 2023.

\bibitem{epoch2023trendsinthedollartrainingcostofmachinelearningsystems}
\BIBentryALTinterwordspacing
B.~Cottier, ``{Trends in the Dollar Training Cost of Machine Learning Systems},'' 2023. [Online]. Available: \url{https://epochai.org/blog/trends-in-the-dollar-training-cost-of-machine-learning-systems}
\BIBentrySTDinterwordspacing

\bibitem{criteo1tb2014}
{Criteo AI Lab}, ``{Criteo 1TB Click Logs Dataset},'' \url{https://ailab.criteo.com/download-criteo-1tb-click-logs-dataset/}, 2014.

\bibitem{dai2018graphh}
G.~Dai, T.~Huang, Y.~Chi, J.~Zhao, G.~Sun, Y.~Liu, Y.~Wang, Y.~Xie, and H.~Yang, ``{GraphH: A Processing-in-Memory Architecture for Large-Scale Graph Processing},'' \emph{{TCAD}}, 2018.

\bibitem{dai2022dimmining}
G.~Dai, Z.~Zhu, T.~Fu, C.~Wei, B.~Wang, X.~Li, Y.~Xie, H.~Yang, and Y.~Wang, ``{DIMMining: Pruning-Efficient and Parallel Graph Mining on Near-Memory-Computing},'' in \emph{{ISCA}}, 2022.

\bibitem{das2022implementation}
P.~Das, P.~R. Sutradhar, M.~Indovina, S.~M.~P. Dinakarrao, and A.~Ganguly, ``{Implementation and Evaluation of Deep Neural Networks in Commercially Available Processing in Memory Hardware},'' in \emph{{SOCC}}, 2022.

\bibitem{de2017understanding}
C.~De~Sa, M.~Feldman, C.~R{\'e}, and K.~Olukotun, ``{Understanding and Optimizing Asynchronous Low-Precision Stochastic Gradient Descent},'' in \emph{{ISCA}}, 2017.

\bibitem{dekel2012optimal}
O.~Dekel, R.~Gilad-Bachrach, O.~Shamir, and L.~Xiao, ``{Optimal Distributed Online Prediction Using Mini-Batches},'' \emph{{JMLR}}, 2012.

\bibitem{deng2018dracc}
Q.~Deng, L.~Jiang, Y.~Zhang, M.~Zhang, and J.~Yang, ``{DrAcc: a DRAM based Accelerator for Accurate CNN Inference},'' in \emph{{DAC}}, 2018.

\bibitem{denzler2023casper}
A.~Denzler, G.~F. Oliveira, N.~Hajinazar, R.~Bera, G.~Singh, J.~G{\'o}mez-Luna, and O.~Mutlu, ``{Casper: Accelerating Stencil Computations Using Near-Cache Processing},'' \emph{IEEE Access}, 2023.

\bibitem{dettmers2022gpt3}
T.~Dettmers, M.~Lewis, Y.~Belkada, and L.~Zettlemoyer, ``{GPT3.int8(): 8-bit Matrix Multiplication for Transformers at Scale},'' in \emph{{NeurIPS}}, 2022.

\bibitem{devaux2019true}
F.~Devaux, ``{The true Processing in Memory accelerator},'' in \emph{{HCS}}, 2019.

\bibitem{diab2023framework}
S.~Diab, A.~Nassereldine, M.~Alser, J.~G{\'o}mez~Luna, O.~Mutlu, and I.~El~Hajj, ``{A Framework for High-throughput Sequence Alignment using Real Processing-in-Memory Systems},'' \emph{{Bioinformatics}}, 2023.

\bibitem{drumond2017mondrian}
M.~Drumond, A.~Daglis, N.~Mirzadeh, D.~Ustiugov, J.~Picorel, B.~Falsafi, B.~Grot, and D.~Pnevmatikatos, ``{The Mondrian Data Engine},'' 2017.

\bibitem{dunner2018snap}
C.~D{\"u}nner, T.~Parnell, D.~Sarigiannis, N.~Ioannou, A.~Anghel, G.~Ravi, M.~Kandasamy, and H.~Pozidis, ``{Snap ML: A Hierarchical Framework for Machine Learning},'' in \emph{{NeurIPS}}, 2018.

\bibitem{falahati2018origami}
H.~Falahati, P.~Lotfi-Kamran, M.~Sadrosadati, and H.~Sarbazi-Azad, ``{ORIGAMI: A Heterogeneous Split Architecture for In-Memory Acceleration of Learning},'' \emph{arXiv:1812.11473}, 2018.

\bibitem{LIBSVMData2023}
R.-E. Fan, ``{LIBSVM Data: A Collection of Benchmarks for Support Vector Machine Research},'' \url{https://www.csie.ntu.edu.tw/~cjlin/libsvmtools/datasets/}.

\bibitem{farmahini2015nda}
A.~Farmahini-Farahani, J.~H. Ahn, K.~Morrow, and N.~S. Kim, ``{NDA: Near-DRAM Acceleration Architecture Leveraging Commodity DRAM Devices and Standard Memory Modules},'' in \emph{{HPCA}}, 2015.

\bibitem{fernandez2024matsa}
I.~Fernandez, C.~Giannoula, A.~Manglik, R.~Quislant, N.~M. Ghiasi, J.~G{\'o}mez-Luna, E.~Gutierrez, O.~Plata, and O.~Mutlu, ``{MATSA: An MRAM-based Energy-Efficient Accelerator for Time Series Analysis},'' \emph{{IEEE Access}}, 2024.

\bibitem{fernandez2020natsa}
I.~Fernandez, R.~Quislant, E.~Guti{\'e}rrez, O.~Plata, C.~Giannoula, M.~Alser, J.~G{\'o}mez-Luna, and O.~Mutlu, ``{NATSA: A Near-Data Processing Accelerator for Time Series Analysis},'' in \emph{{ICCD}}, 2020.

\bibitem{ferreira2022pluto}
J.~D. Ferreira, G.~Falcao, J.~G{\'o}mez-Luna, M.~Alser, L.~Orosa, M.~Sadrosadati, J.~S. Kim, G.~F. Oliveira, T.~Shahroodi, A.~Nori \emph{et~al.}, ``{pLUTo: Enabling Massively Parallel Computation in DRAM via Lookup Tables},'' in \emph{{MICRO}}, 2022.

\bibitem{gao2019computedram}
F.~Gao, G.~Tziantzioulis, and D.~Wentzlaff, ``{ComputeDRAM: In-Memory Compute Using Off-the-Shelf DRAMs},'' in \emph{{MICRO}}, 2019.

\bibitem{gao2015practical}
M.~Gao, G.~Ayers, and C.~Kozyrakis, ``{Practical Near-Data Processing for In-Memory Analytics Frameworks},'' in \emph{{PACT}}, 2015.

\bibitem{gao2016hrl}
M.~Gao and C.~Kozyrakis, ``{HRL: Efficient and flexible reconfigurable logic for near-data processing},'' in \emph{{HPCA}}, 2016.

\bibitem{gao2017tetris}
M.~Gao, J.~Pu, X.~Yang, M.~Horowitz, and C.~Kozyrakis, ``{TETRIS: Scalable and Efficient Neural Network Acceleration with 3D Memory},'' in \emph{{ASPLOS}}, 2017.

\bibitem{geman1992neural}
S.~Geman, E.~Bienenstock, and R.~Doursat, ``{Neural Networks and the Bias/Variance Dilemma},'' \emph{{Neural Computation}}, 1992.

\bibitem{ghiasi2024megis}
N.~M. Ghiasi, M.~Sadrosadati, H.~Mustafa, A.~Gollwitzer, C.~Firtina, J.~Eudine, H.~Mao, J.~Lindegger, M.~B. Cavlak, M.~Alser, J.~Park, and O.~Mutlu, ``{MegIS: High-Performance, Energy-Efficient, and Low-Cost Metagenomic Analysis with In-Storage Processing},'' in \emph{{ISCA}}, 2024.

\bibitem{ghiasi2022alp}
N.~M. Ghiasi, N.~Vijaykumar, G.~F. Oliveira, L.~Orosa, I.~Fernandez, M.~Sadrosadati, K.~Kanellopoulos, N.~Hajinazar, J.~G. Luna, and O.~Mutlu, ``{ALP: Alleviating CPU-Memory Data Movement Overheads in Memory-Centric Systems},'' \emph{{IEEE Transactions on Emerging Topics in Computing}}, 2022.

\bibitem{ghose2019processing}
S.~Ghose, A.~Boroumand, J.~S. Kim, J.~G{\'o}mez-Luna, and O.~Mutlu, ``{Processing-In-Memory: A Workload-Driven Perspective},'' \emph{{IBM Journal of Research and Development}}, 2019.

\bibitem{giannoula2022sparsep}
C.~Giannoula, I.~Fernandez, J.~G. Luna, N.~Koziris, G.~Goumas, and O.~Mutlu, ``{SparseP: Towards Efficient Sparse Matrix Vector Multiplication on Real Processing-In-Memory Architectures},'' \emph{{POMACS}}, 2022.

\bibitem{giannoula2021syncron}
C.~Giannoula, N.~Vijaykumar, N.~Papadopoulou, V.~Karakostas, I.~Fernandez, J.~G{\'o}mez-Luna, L.~Orosa, N.~Koziris, G.~Goumas, and O.~Mutlu, ``{SynCron: Efficient Synchronization Support for Near-Data-Processing Architectures},'' in \emph{{HPCA}}, 2021.

\bibitem{giannoula2024accelerating}
C.~Giannoula, P.~Yang, I.~F. Vega, J.~Yang, Y.~X. Li, J.~G. Luna, M.~Sadrosadati, O.~Mutlu, and G.~Pekhimenko, ``{Accelerating Graph Neural Networks on Real Processing-In-Memory Systems},'' \emph{arXiv:2402.16731}, 2024.

\bibitem{gogineni2024swiftrl}
K.~Gogineni, S.~S. Dayapule, J.~G{\'o}mez-Luna, K.~Gogineni, P.~Wei, T.~Lan, M.~Sadrosadati, O.~Mutlu, and G.~Venkataramani, ``{SwiftRL: Towards Efficient Reinforcement Learning on Real Processing-In-Memory Systems},'' \emph{arXiv:2405.03967}, 2024.

\bibitem{gomez2021benchmarking}
J.~G{\'o}mez-Luna, I.~El~Hajj, I.~Fernandez, C.~Giannoula, G.~F. Oliveira, and O.~Mutlu, ``{Benchmarking Memory-Centric Computing Systems: Analysis of Real Processing-In-Memory Hardware},'' in \emph{{IGSC}}, 2021.

\bibitem{gomez2022benchmarking}
------, ``{Benchmarking a New Paradigm: Experimental Analysis and Characterization of a Real Processing-In-Memory System},'' \emph{{IEEE Access}}, 2022.

\bibitem{gomez2022experimental}
J.~G{\'o}mez-Luna, Y.~Guo, S.~Brocard, J.~Legriel, R.~Cimadomo, G.~F. Oliveira, G.~Singh, and O.~Mutlu, ``{An Experimental Evaluation of Machine Learning Training on a Real Processing-In-Memory System},'' \emph{arXiv:2207.07886}, 2022.

\bibitem{gomez2023evaluatingispass}
------, ``{Evaluating Machine Learning Workloads on Memory-Centric Computing Systems},'' in \emph{{ISPASS}}, 2023.

\bibitem{goodfellow2016}
I.~Goodfellow, Y.~Bengio, and A.~Courville, \emph{{Deep Learning}}.\hskip 1em plus 0.5em minus 0.4em\relax {The MIT Press}, 2016.

\bibitem{gu2016biscuit}
B.~Gu, A.~S. Yoon, D.-H. Bae, I.~Jo, J.~Lee, J.~Yoon, J.-U. Kang, M.~Kwon, C.~Yoon, S.~Cho \emph{et~al.}, ``{Biscuit: A Framework for Near-Data Processing of Big Data Workloads},'' in \emph{{ISCA}}, 2016.

\bibitem{gupta2023evaluating}
H.~Gupta, M.~Kabra, J.~G{\'o}mez-Luna, K.~Kanellopoulos, and O.~Mutlu, ``{Evaluating Homomorphic Operations on a Real-World Processing-In-Memory System},'' in \emph{{IISWC}}, 2023.

\bibitem{gomezluna2021prim}
J.~Gómez-Luna, I.~E. Hajj, I.~Fernandez, C.~Giannoula, G.~F. Oliveira, and O.~Mutlu, ``{Benchmarking a New Paradigm: An Experimental Analysis of a Real Processing-in-Memory Architecture},'' \emph{arXiv:2105.03814}, 2021.

\bibitem{hajinazar2021simdram}
N.~Hajinazar, G.~F. Oliveira, S.~Gregorio, J.~D. Ferreira, N.~M. Ghiasi, M.~Patel, M.~Alser, S.~Ghose, J.~G{\'o}mez-Luna, and O.~Mutlu, ``{SIMDRAM: An End-to-End Framework for Bit-Serial SIMD Computing in DRAM},'' in \emph{{ASPLOS}}, 2021.

\bibitem{han2015deep}
S.~Han, H.~Mao, and W.~J. Dally, ``{Deep Compression: Compressing Deep Neural Networks with Pruning, Trained Quantization and Huffman Coding},'' \emph{arXiv:1510.00149}, 2015.

\bibitem{hashemi2016accelerating}
M.~Hashemi, Khubaib, E.~Ebrahimi, O.~Mutlu, and Y.~N. Patt, ``{Accelerating Dependent Cache Misses with an Enhanced Memory Controller},'' in \emph{{ISCA}}, 2016.

\bibitem{hashemi2016continuous}
M.~Hashemi, O.~Mutlu, and Y.~N. Patt, ``{Continuous Runahead: Transparent Hardware Acceleration for Memory Intensive Workloads},'' in \emph{{MICRO}}, 2016.

\bibitem{hassan2015near}
S.~M. Hassan, S.~Yalamanchili, and S.~Mukhopadhyay, ``{Near Data Processing: Impact and Optimization of 3D Memory System Architecture on the Uncore},'' in \emph{{MEMSYS}}, 2015.

\bibitem{hastie2009elements}
T.~Hastie, R.~Tibshirani, and J.~Friedman, \emph{{The Elements of Statistical Learning}}.\hskip 1em plus 0.5em minus 0.4em\relax Springer, 2009.

\bibitem{herruzo2021enabling}
J.~M. Herruzo, I.~Fernandez, S.~Gonz{\'a}lez-Navarro, and O.~Plata, ``{Enabling Fast and Energy-Efficient FM-index Exact Matching using Processing-Near-Memory},'' \emph{{The Journal of Supercomputing}}, 2021.

\bibitem{hsieh2016transparent}
K.~Hsieh, E.~Ebrahimi, G.~Kim, N.~Chatterjee, M.~O'Connor, N.~Vijaykumar, O.~Mutlu, and S.~W. Keckler, ``{Transparent Offloading and Mapping (TOM): Enabling Programmer-Transparent Near-Data Processing in GPU Systems},'' in \emph{{ISCA}}, 2016.

\bibitem{hsieh2016accelerating}
K.~Hsieh, S.~Khan, N.~Vijaykumar, K.~K. Chang, A.~Boroumand, S.~Ghose, and O.~Mutlu, ``{Accelerating Pointer Chasing in 3D-Stacked Memory: Challenges, Mechanisms, Evaluation},'' in \emph{{ICCD}}, 2016.

\bibitem{huang2005using}
J.~Huang and C.~X. Ling, ``{Using AUC and Accuracy in Evaluating Learning Algorithms},'' \emph{{IEEE Transactions on Knowledge and Data Engineering}}, 2005.

\bibitem{hyun2023pathfinding}
B.~Hyun, T.~Kim, D.~Lee, and M.~Rhu, ``{Pathfinding Future PIM Architectures by Demystifying a Commercial PIM Technology},'' \emph{arXiv:2308.00846}, 2023.

\bibitem{hyun2024pathfinding}
------, ``{Pathfinding Future PIM Architectures by Demystifying a Commercial PIM Technology},'' in \emph{{HPCA}}, 2024.

\bibitem{imani2019floatpim}
M.~Imani, S.~Gupta, Y.~Kim, and T.~Rosing, ``{FloatPIM: In-Memory Acceleration of Deep Neural Network Training with High Precision},'' in \emph{{ISCA}}, 2019.

\bibitem{10158230}
M.~Item, J.~Gómez-Luna, G.~F. Oliveira, M.~Sadrosadati, Y.~Guo, and O.~Mutlu, ``{TransPimLib: Efficient Transcendental Functions for Processing-in-Memory Systems},'' in \emph{{ISPASS}}, 2023.

\bibitem{ivanov2021data}
A.~Ivanov, N.~Dryden, T.~Ben-Nun, S.~Li, and T.~Hoefler, ``{Data Movement is All You Need: A Case Study on Optimizing Transformers},'' in \emph{{MLSys}}, 2021.

\bibitem{jiang2019cimat}
H.~Jiang, X.~Peng, S.~Huang, and S.~Yu, ``{CIMAT: A Transpose SRAM-based Compute-In-Memory Architecture for Deep Neural Network On-Chip Training},'' in \emph{{MEMSYS}}, 2019.

\bibitem{jiang2021towards}
J.~Jiang, S.~Gan, Y.~Liu, F.~Wang, G.~Alonso, A.~Klimovic, A.~Singla, W.~Wu, and C.~Zhang, ``{Towards Demystifying Serverless Machine Learning Training},'' in \emph{SIGMOD}, 2021.

\bibitem{jonatan2024scalability}
G.~Jonatan, H.~Cho, H.~Son, X.~Wu, N.~Livesay, E.~Mora, K.~Shivdikar, J.~L. Abell{\'a}n, A.~Joshi, D.~Kaeli \emph{et~al.}, ``{Scalability Limitations of Processing-in-Memory using Real System Evaluations},'' \emph{{POMACS}}, 2024.

\bibitem{kang2023pim}
H.~Kang, Y.~Zhao, G.~E. Blelloch, L.~Dhulipala, Y.~Gu, C.~McGuffey, and P.~B. Gibbons, ``{PIM-trie: A Skew-resistant Trie for Processing-in-Memory},'' in \emph{{SPAA}}, 2023.

\bibitem{kautz1969cellular}
W.~H. Kautz, ``{Cellular Logic-in-Memory Arrays},'' \emph{{IEEE Transactions on Computers}}, 1969.

\bibitem{ke2020recnmp}
L.~Ke, U.~Gupta, B.~Y. Cho, D.~Brooks, V.~Chandra, U.~Diril, A.~Firoozshahian, K.~Hazelwood, B.~Jia, H.-H.~S. Lee \emph{et~al.}, ``{RecNMP: Accelerating Personalized Recommendation with Near-Memory Processing},'' in \emph{{ISCA}}, 2020.

\bibitem{ke2021near}
L.~Ke, X.~Zhang, J.~So, J.-G. Lee, S.-H. Kang, S.~Lee, S.~Han, Y.~Cho, J.~H. Kim, Y.~Kwon \emph{et~al.}, ``{Near-Memory Processing in Action: Accelerating Personalized Recommendation With AXDIMM},'' \emph{{MICRO}}, 2021.

\bibitem{khan2024landscape}
A.~A. Khan, J.~P.~C. De~Lima, H.~Farzaneh, and J.~Castrillon, ``{The Landscape of Compute-Near-Memory and Compute-In-Memory: A Research and Commercial Overview},'' \emph{arXiv:2401.14428}, 2024.

\bibitem{khan2022cinm}
A.~A. Khan, H.~Farzaneh, K.~F. Friebel, C.~Fournier, L.~Chelini, and J.~Castrillon, ``{CINM (Cinnamon): A Compilation Infrastructure for Heterogeneous Compute In-Memory and Compute Near-Memory Paradigms},'' \emph{arXiv:2301.07486}, 2022.

\bibitem{kim2020mvid}
B.~Kim, J.~Chung, E.~Lee, W.~Jung, S.~Lee, J.~Choi, J.~Park, M.~Wi, S.~Lee, and J.~H. Ahn, ``{MViD: Sparse Matrix-Vector Multiplication in Mobile DRAM for Accelerating Recurrent Neural Networks},'' \emph{{IEEE Transactions on Computers}}, 2020.

\bibitem{kim2016neurocube}
D.~Kim, J.~Kung, S.~Chai, S.~Yalamanchili, and S.~Mukhopadhyay, ``{Neurocube: A Programmable Digital Neuromorphic Architecture with High-Density 3D Memory},'' in \emph{{ISCA}}, 2016.

\bibitem{kim2021gradpim}
H.~Kim, H.~Park, T.~Kim, K.~Cho, E.~Lee, S.~Ryu, H.-J. Lee, K.~Choi, and J.~Lee, ``{GradPIM: A Practical Processing-in-DRAM Architecture for Gradient Descent},'' in \emph{{HPCA}}, 2021.

\bibitem{kim2017grim}
J.~S. Kim, D.~Senol, H.~Xin, D.~Lee, S.~Ghose, M.~Alser, H.~Hassan, O.~Ergin, C.~Alkan, and O.~Mutlu, ``{GRIM-Filter: Fast Seed Location Filtering in DNA Read Mapping Using Processing-in-Memory Technologies},'' \emph{arXiv:1708.04329}, 2017.

\bibitem{kim2024optimal}
S.~Y. Kim, J.~Lee, Y.~Paik, C.~H. Kim, W.~J. Lee, and S.~W. Kim, ``{Optimal Model Partitioning with Low-Overhead Profiling on the PIM-based Platform for Deep Learning Inference},'' \emph{{TODAES}}, 2024.

\bibitem{kingma2014adam}
D.~P. Kingma, ``{Adam: A Method for Stochastic Optimization},'' \emph{arXiv:1412.6980}, 2014.

\bibitem{kwon202125}
Y.-C. Kwon, S.~H. Lee, J.~Lee, S.-H. Kwon, J.~M. Ryu, J.-P. Son, O.~Seongil, H.-S. Yu, H.~Lee, S.~Y. Kim \emph{et~al.}, ``{25.4 A 20nm 6GB Function-In-Memory DRAM, Based on HBM2 with a 1.2TFLOPS Programmable Computing Unit Using Bank-Level Parallelism, for Machine Learning Applications},'' in \emph{{ISSCC}}, 2021.

\bibitem{kwon2019tensordimm}
Y.~Kwon, Y.~Lee, and M.~Rhu, ``{TensorDIMM: A Practical Near-Memory Processing Architecture for Embeddings and Tensor Operations in Deep Learning},'' in \emph{{MICRO}}, 2019.

\bibitem{lan2020communication}
G.~Lan, S.~Lee, and Y.~Zhou, ``{Communication-Efficient Algorithms for Decentralized and Stochastic Optimization},'' \emph{{Mathematical Programming}}, 2020.

\bibitem{lavenier2020variant}
D.~Lavenier, R.~Cimadomo, and R.~Jodin, ``{Variant Calling Parallelization on Processor-in-Memory Architecture},'' in \emph{{BIBM}}, 2020.

\bibitem{lavenier2016blast}
D.~Lavenier, C.~Deltel, D.~Furodet, and J.-F. Roy, ``{BLAST on UPMEM},'' Ph.D. dissertation, {INRIA Rennes-Bretagne Atlantique}, 2016.

\bibitem{lee2022improving}
D.~Lee, J.~So, M.~Ahn, J.-G. Lee, J.~Kim, J.~Cho, R.~Oliver, V.~C. Thummala, R.~s. JV, S.~S. Upadhya \emph{et~al.}, ``{Improving In-Memory Database Operations with Acceleration DIMM (AxDIMM)},'' in \emph{{DaMoN}}, 2022.

\bibitem{lee2015bssync}
J.~H. Lee, J.~Sim, and H.~Kim, ``{BSSync: Processing Near Memory for Machine Learning Workloads with Bounded Staleness Consistency Models},'' in \emph{{PACT}}, 2015.

\bibitem{lee20221ynm}
S.~Lee, K.~Kim, S.~Oh, J.~Park, G.~Hong, D.~Ka, K.~Hwang, J.~Park, K.~Kang, J.~Kim \emph{et~al.}, ``{A 1ynm 1.25V 8Gb, 16Gb/s/pin GDDR6-based Accelerator-in-Memory supporting 1TFLOPS MAC Operation and Various Activation Functions for Deep-Learning Applications},'' in \emph{{ISSCC}}, 2022.

\bibitem{lee2021hardware}
S.~Lee, S.-h. Kang, J.~Lee, H.~Kim, E.~Lee, S.~Seo, H.~Yoon, S.~Lee, K.~Lim, H.~Shin \emph{et~al.}, ``{Hardware Architecture and Software Stack for PIM Based on Commercial DRAM Technology},'' in \emph{{ISCA}}, 2021.

\bibitem{lee2021task}
Y.~S. Lee and T.~H. Han, ``{Task Parallelism-Aware Deep Neural Network Scheduling on Multiple Hybrid Memory Cube-based Processing-in-Memory},'' \emph{{IEEE Access}}, 2021.

\bibitem{li20203d}
B.~Li, J.~R. Doppa, P.~P. Pande, K.~Chakrabarty, J.~X. Qiu, and H.~Li, ``{3D-ReG: A 3D ReRAM-based Heterogeneous Architecture for Training Deep Neural Networks},'' \emph{{JETC}}, 2020.

\bibitem{li2014communication}
M.~Li, D.~G. Andersen, A.~J. Smola, and K.~Yu, ``{Communication Efficient Distributed Machine Learning with the Parameter Server},'' 2014.

\bibitem{li2014efficient}
M.~Li, T.~Zhang, Y.~Chen, and A.~J. Smola, ``{Efficient Mini-Batch Training for Stochastic Optimization},'' in \emph{{SIGKDD}}, 2014.

\bibitem{lian2017can}
X.~Lian, C.~Zhang, H.~Zhang, C.-J. Hsieh, W.~Zhang, and J.~Liu, ``{Can Decentralized Algorithms Outperform Centralized Algorithms? A Case Study for Decentralized Parallel Stochastic Gradient Descent},'' in \emph{{NeurIPS}}, 2017.

\bibitem{lim2023design}
C.~Lim, S.~Lee, J.~Choi, J.~Lee, S.~Park, H.~Kim, J.~Lee, and Y.~Kim, ``{Design and Analysis of a Processing-in-DIMM Join Algorithm: A Case Study with UPMEM DIMMs},'' \emph{{PACMMOD}}, 2023.

\bibitem{liu2020distributed}
J.~Liu, C.~Zhang \emph{et~al.}, ``{Distributed Learning Systems with First-Order Methods},'' \emph{{Foundations and Trends{\textregistered} in Databases}}, 2020.

\bibitem{liu2018processing}
J.~Liu, H.~Zhao, M.~A. Ogleari, D.~Li, and J.~Zhao, ``{Processing-In-Memory for Energy-Efficient Neural Network Training: A Heterogeneous Approach},'' in \emph{{MICRO}}, 2018.

\bibitem{liu2017concurrent}
Z.~Liu, I.~Calciu, M.~Herlihy, and O.~Mutlu, ``{Concurrent Data Structures for Near-Memory Computing},'' in \emph{{SPAA}}, 2017.

\bibitem{luo2020benchmark}
Y.~Luo and S.~Yu, ``{Benchmark Non-Volatile and Volatile Memory Based Hybrid Precision Synapses for In-Situ Deep Neural Network Training},'' in \emph{{ASP-DAC}}, 2020.

\bibitem{mahajan2016tabla}
D.~Mahajan, J.~Park, E.~Amaro, H.~Sharma, A.~Yazdanbakhsh, J.~K. Kim, and H.~Esmaeilzadeh, ``{TABLA: A Unified Template-based Framework for Accelerating Statistical Machine Learning},'' in \emph{{HPCA}}, 2016.

\bibitem{mansouri2022genstore}
N.~Mansouri~Ghiasi, J.~Park, H.~Mustafa, J.~Kim, A.~Olgun, A.~Gollwitzer, D.~Senol~Cali, C.~Firtina, H.~Mao, N.~Almadhoun~Alserr \emph{et~al.}, ``{GenStore: A High-Performance In-Storage Processing System for Genome Sequence Analysis},'' in \emph{{ASPLOS}}, 2022.

\bibitem{mao2022genpip}
H.~Mao, M.~Alser, M.~Sadrosadati, C.~Firtina, A.~Baranwal, D.~S. Cali, A.~Manglik, N.~A. Alserr, and O.~Mutlu, ``{GenPIP: In-Memory Acceleration of Genome Analysis via Tight Integration of Basecalling and Read Mapping},'' in \emph{{MICRO}}, 2022.

\bibitem{mcdonald2010distributed}
R.~McDonald, K.~Hall, and G.~Mann, ``{Distributed Training Strategies for the Structured Perceptron},'' in \emph{{NAACL HLT}}, 2010.

\bibitem{mutlu2020intelligent}
O.~Mutlu, ``{Intelligent Architectures for Intelligent Computing Systems},'' \emph{DATE}, 2021.

\bibitem{mutlu2023dac}
------, ``{Memory-Centric Computing},'' in \emph{{DAC}}, 2023.

\bibitem{mutlu2019enabling}
O.~Mutlu, S.~Ghose, J.~G{\'o}mez-Luna, and R.~Ausavarungnirun, ``{Enabling Practical Processing in and near Memory for Data-Intensive Computing},'' in \emph{{DAC}}, 2019.

\bibitem{mutlu2019processing}
------, ``{Processing Data Where It Makes Sense: Enabling In-Memory Computation},'' \emph{{Microprocessors and Microsystems}}, 2019.

\bibitem{mutlu2022modern}
------, ``{A Modern Primer on Processing in Memory},'' in \emph{{Emerging Computing: From Devices to Systems: Looking Beyond Moore and Von Neumann}}.\hskip 1em plus 0.5em minus 0.4em\relax Springer, 2022.

\bibitem{nai2017graphpim}
L.~Nai, R.~Hadidi, J.~Sim, H.~Kim, P.~Kumar, and H.~Kim, ``{GraphPIM: Enabling Instruction-Level PIM Offloading in Graph Computing Frameworks},'' in \emph{{HPCA}}, 2017.

\bibitem{nider2021case}
J.~Nider, C.~Mustard, A.~Zoltan, J.~Ramsden, L.~Liu, J.~Grossbard, M.~Dashti, R.~Jodin, A.~Ghiti, J.~Chauzi \emph{et~al.}, ``{A Case Study of Processing-in-Memory in off-the-Shelf Systems},'' in \emph{{USENIX}}, 2021.

\bibitem{niu2022184qps}
D.~Niu, S.~Li, Y.~Wang, W.~Han, Z.~Zhang, Y.~Guan, T.~Guan, F.~Sun, F.~Xue, L.~Duan \emph{et~al.}, ``{184QPS/W 64Mb/mm2 3D Logic-to-DRAM Hybrid Bonding with Process-Near-Memory Engine for Recommendation System},'' in \emph{{ISSCC}}, 2022.

\bibitem{nvidia2020a100}
NVIDIA, ``{NVIDIA A100 Tensor Core GPU Architecture. White Paper},'' \url{https://images.nvidia.com/aem-dam/en-zz/Solutions/data-center/nvidia-ampere-architecture-whitepaper.pdf}, 2020.

\bibitem{olgun2022pidram}
A.~Olgun, J.~G. Luna, K.~Kanellopoulos, B.~Salami, H.~Hassan, O.~Ergin, and O.~Mutlu, ``{PiDRAM: A Holistic End-to-end FPGA-based Framework for Processing-in-DRAM},'' \emph{{TACO}}, 2022.

\bibitem{oliveira2022heterogeneous}
G.~F. Oliveira, A.~Boroumand, S.~Ghose, J.~G{\'o}mez-Luna, and O.~Mutlu, ``{Heterogeneous Data-Centric Architectures for Modern Data-Intensive Applications: Case Studies in Machine Learning and Databases},'' in \emph{{ISVLSI}}, 2022.

\bibitem{oliveira2021damov}
G.~F. Oliveira, J.~G{\'o}mez-Luna, L.~Orosa, S.~Ghose, N.~Vijaykumar, I.~Fernandez, M.~Sadrosadati, and O.~Mutlu, ``{DAMOV: A New Methodology and Benchmark Suite for Evaluating Data Movement Bottlenecks},'' \emph{{IEEE Access}}, 2021.

\bibitem{oliveira2023dappa}
G.~F. Oliveira, A.~Kohli, D.~Novo, J.~G{\'o}mez-Luna, and O.~Mutlu, ``{DaPPA: A Data-Parallel Framework for Processing-In-Memory Architectures},'' \emph{arXiv:2310.10168}, 2023.

\bibitem{oliveira2024mimdram}
G.~F. Oliveira, A.~Olgun, A.~G. Ya{\u{g}}l{\i}k{\c{c}}{\i}, F.~N. Bostanc{\i}, J.~G{\'o}mez-Luna, S.~Ghose, and O.~Mutlu, ``{MIMDRAM: An End-to-End Processing-Using-DRAM System for High-Throughput, Energy-Efficient and Programmer-Transparent Multiple-Instruction Multiple-Data Computing},'' in \emph{{HPCA}}, 2024.

\bibitem{park2021trim}
J.~Park, B.~Kim, S.~Yun, E.~Lee, M.~Rhu, and J.~H. Ahn, ``{TRiM: Enhancing Processor-Memory Interfaces with Scalable Tensor Reduction in Memory},'' in \emph{{MICRO}}, 2021.

\bibitem{park2022flash}
J.~Park, R.~Azizi, G.~F. Oliveira, M.~Sadrosadati, R.~Nadig, D.~Novo, J.~G{\'o}mez-Luna, M.~Kim, and O.~Mutlu, ``{Flash-Cosmos: In-Flash Bulk Bitwise Operations Using Inherent Computation Capability of NAND Flash Memory},'' in \emph{{MICRO}}, 2022.

\bibitem{park2021high}
N.~Park, S.~Ryu, J.~Kung, and J.-J. Kim, ``{High-throughput Near-Memory Processing on CNNs with 3D HBM-like Memory},'' \emph{{TODAES}}, 2021.

\bibitem{paszke2019pytorch}
A.~Paszke, S.~Gross, F.~Massa, A.~Lerer, J.~Bradbury, G.~Chanan, T.~Killeen, Z.~Lin, N.~Gimelshein, L.~Antiga \emph{et~al.}, ``{PyTorch: An Imperative Style, High-Performance Deep Learning Library},'' in \emph{{NeurIPS}}, 2019.

\bibitem{pattnaik2016scheduling}
A.~Pattnaik, X.~Tang, A.~Jog, O.~Kayiran, A.~K. Mishra, M.~T. Kandemir, O.~Mutlu, and C.~R. Das, ``{Scheduling Techniques for GPU Architectures with Processing-In-Memory Capabilities},'' in \emph{{PACT}}, 2016.

\bibitem{peccerillo2022survey}
B.~Peccerillo, M.~Mannino, A.~Mondelli, and S.~Bartolini, ``{A Survey on Hardware Accelerators: Taxonomy, Trends, Challenges, and Perspectives},'' \emph{{Journal of Systems Architecture}}, 2022.

\bibitem{polyak1987introduction}
B.~T. Polyak, \emph{{Introduction to Optimization}}.\hskip 1em plus 0.5em minus 0.4em\relax {Optimization Software}, 1987.

\bibitem{pugsley2014ndc}
S.~H. Pugsley, J.~Jestes, H.~Zhang, R.~Balasubramonian, V.~Srinivasan, A.~Buyuktosunoglu, A.~Davis, and F.~Li, ``{NDC: Analyzing the Impact of 3D-Stacked Memory+Logic Devices on MapReduce Workloads},'' in \emph{{ISPASS}}, 2014.

\bibitem{robbins1951stochastic}
H.~Robbins and S.~Monro, ``{A Stochastic Approximation Method},'' \emph{{The Annals of Mathematical Statistics}}, 1951.

\bibitem{ruder2016overview}
S.~Ruder, ``{An Overview of Gradient Descent Optimization Algorithms},'' \emph{arXiv:1609.04747}, 2016.

\bibitem{pimopt-artifact-github}
{SAFARI Research Group}, ``{PIM-Opt Artifact --- GitHub Repository},'' \url{https://github.com/CMU-SAFARI/PIM-Opt}, 2024.

\bibitem{zenodo-artifact}
------, ``{PIM-Opt Artifact --- Zenodo Repository},'' \url{https://doi.org/10.5281/zenodo.12747665}, 2024.

\bibitem{saikia2019k}
J.~Saikia, S.~Yin, Z.~Jiang, M.~Seok, and J.-s. Seo, ``{K-Nearest Neighbor Hardware Accelerator Using In-Memory Computing SRAM},'' in \emph{{ISLPED}}, 2019.

\bibitem{schuiki2018scalable}
F.~Schuiki, M.~Schaffner, F.~K. G{\"u}rkaynak, and L.~Benini, ``{A Scalable Near-Memory Architecture for Training Deep Neural Networks on Large In-Memory Datasets},'' \emph{{IEEE Transactions on Computers}}, 2018.

\bibitem{seshadri2016simple}
V.~Seshadri, ``{Simple DRAM and Virtual Memory Abstractions to Enable Highly Efficient Memory Systems},'' \emph{arXiv:1605.06483}, 2016.

\bibitem{seshadri2015fast}
V.~Seshadri, K.~Hsieh, A.~Boroum, D.~Lee, M.~A. Kozuch, O.~Mutlu, P.~B. Gibbons, and T.~C. Mowry, ``{Fast Bulk Bitwise AND and OR in DRAM},'' \emph{{ICAL}}, 2015.

\bibitem{seshadri2013rowclone}
V.~Seshadri, Y.~Kim, C.~Fallin, D.~Lee, R.~Ausavarungnirun, G.~Pekhimenko, Y.~Luo, O.~Mutlu, P.~B. Gibbons, M.~A. Kozuch \emph{et~al.}, ``{RowClone: Fast and Energy-Efficient In-DRAM Bulk Data Copy and Initialization},'' in \emph{{MICRO}}, 2013.

\bibitem{seshadri2016buddy}
V.~Seshadri, D.~Lee, T.~Mullins, H.~Hassan, A.~Boroumand, J.~Kim, M.~A. Kozuch, O.~Mutlu, P.~B. Gibbons, and T.~C. Mowry, ``{Buddy-RAM: Improving the Performance and Efficiency of Bulk Bitwise Operations Using DRAM},'' \emph{arXiv:1611.09988}, 2016.

\bibitem{seshadri2017ambit}
------, ``{Ambit: In-Memory Accelerator for Bulk Bitwise Operations Using Commodity DRAM Technology},'' in \emph{{MICRO}}, 2017.

\bibitem{seshadri2017simple}
V.~Seshadri and O.~Mutlu, ``{Simple Operations in Memory to Reduce Data Movement},'' in \emph{{Advances in Computers}}.\hskip 1em plus 0.5em minus 0.4em\relax Elsevier, 2017.

\bibitem{seshadri2019dram}
------, ``{In-DRAM Bulk Bitwise Execution Engine},'' \emph{arXiv:1905.09822}, 2019.

\bibitem{shahroodi2023swordfish}
T.~Shahroodi, G.~Singh, M.~Zahedi, H.~Mao, J.~Lindegger, C.~Firtina, S.~Wong, O.~Mutlu, and S.~Hamdioui, ``{Swordfish: A Framework for Evaluating Deep Neural Network-based Basecalling using Computation-In-Memory with Non-Ideal Memristors},'' in \emph{{MICRO}}, 2023.

\bibitem{shelor2019reconfigurable}
C.~F. Shelor and K.~M. Kavi, ``{Reconfigurable Dataflow Graphs for Processing-In-Memory},'' in \emph{{ICDCN}}, 2019.

\bibitem{shin2018mcdram}
H.~Shin, D.~Kim, E.~Park, S.~Park, Y.~Park, and S.~Yoo, ``{McDRAM: Low Latency and Energy-Efficient Matrix Computations in DRAM},'' \emph{{TCAD}}, 2018.

\bibitem{singh2020nero}
G.~Singh, D.~Diamantopoulos, C.~Hagleitner, J.~Gomez-Luna, S.~Stuijk, O.~Mutlu, and H.~Corporaal, ``{NERO: A Near High-Bandwidth Memory Stencil Accelerator for Weather Prediction Modeling},'' in \emph{{FPL}}, 2020.

\bibitem{singh2019napel}
G.~Singh, J.~G{\'o}mez-Luna, G.~Mariani, G.~F. Oliveira, S.~Corda, S.~Stuijk, O.~Mutlu, and H.~Corporaal, ``{NAPEL: Near-Memory Computing Application Performance Prediction via Ensemble Learning},'' in \emph{{DAC}}, 2019.

\bibitem{sirb2016consensus}
B.~Sirb and X.~Ye, ``{Consensus Optimization with Delayed and Stochastic Gradients on Decentralized Networks},'' in \emph{{Big Data}}, 2016.

\bibitem{stich2018sparsified}
S.~U. Stich, J.-B. Cordonnier, and M.~Jaggi, ``{Sparsified SGD with Memory},'' in \emph{{NeurIPS}}, 2018.

\bibitem{stone1970logic}
H.~S. Stone, ``{A Logic-in-Memory Computer},'' \emph{{IEEE Transactions on Computers}}, 1970.

\bibitem{strom2015scalable}
N.~Str{\"o}m, ``{Scalable Distributed DNN Training using Commodity GPU Cloud Computing},'' \emph{{Sixteenth Annual Conference of the International Speech Communication Association}}, 2015.

\bibitem{sun2020energy}
H.~Sun, Z.~Zhu, Y.~Cai, X.~Chen, Y.~Wang, and H.~Yang, ``{An Energy-Efficient Quantized and Regularized Training Framework for Processing-In-Memory Accelerators},'' in \emph{{ASP-DAC}}, 2020.

\bibitem{sun2021abc}
W.~Sun, Z.~Li, S.~Yin, S.~Wei, and L.~Liu, ``{ABC-DIMM: Alleviating the Bottleneck of Communication in DIMM-based Near-Memory Processing with Inter-DIMM Broadcast},'' in \emph{{ISCA}}, 2021.

\bibitem{sun2020one}
Z.~Sun, G.~Pedretti, A.~Bricalli, and D.~Ielmini, ``{One-Step Regression and Classification with Cross-Point Resistive Memory Arrays},'' \emph{{Science Advances}}, 2020.

\bibitem{yfcc100m}
B.~Thomee, D.~A. Shamma, G.~Friedland, B.~Elizalde, K.~Ni, D.~Poland, D.~Borth, and L.-J. Li, ``{YFCC100M: The New Data in Multimedia Research},'' \emph{{Communications of the ACM}}, 2016.

\bibitem{umuroglu2017finn}
Y.~Umuroglu, N.~J. Fraser, G.~Gambardella, M.~Blott, P.~Leong, M.~Jahre, and K.~Vissers, ``{FINN: A Framework for Fast, Scalable Binarized Neural Network Inference},'' in \emph{{FPGA}}, 2017.

\bibitem{upmemproductsheet2022}
UPMEM, ``{Product Sheet UPMEM},'' 2022.

\bibitem{upmemtechpaper2022}
{UPMEM}, ``{UPMEM Processing In-Memory (PIM)},'' {UPMEM PIM Tech Paper}, 2022.

\bibitem{abumpimp2023}
UPMEM, ``{UPMEM PIM Platform for Data-Intensive Applications},'' in \emph{{ABUMPIMP}}.\hskip 1em plus 0.5em minus 0.4em\relax {Symposium as part of Euro-Par}, 2023.

\bibitem{upmem_sdk}
------, ``{UPMEM SDK, Version 2023.2.0},'' \url{https://sdk.upmem.com/2023.2.0/}, 2023.

\bibitem{upmemwebsite2024}
------, ``{UPMEM Website},'' \url{https://www.upmem.com}, 2024.

\bibitem{vieira2018exploiting}
J.~Vieira, N.~Roma, P.~Tom{\'a}s, P.~Ienne, and G.~Falcao, ``{Exploiting Compute Caches for Memory Bound Vector Operations},'' in \emph{{SBAC-PAD}}, 2018.

\bibitem{villalobos2022will}
P.~Villalobos, J.~Sevilla, L.~Heim, T.~Besiroglu, M.~Hobbhahn, and A.~Ho, ``{Will we run out of Data? An Analysis of the Limits of scaling datasets in Machine Learning},'' \emph{arXiv:2211.04325}, 2022.

\bibitem{wang2017memory}
J.~Wang, W.~Wang, and N.~Srebro, ``{Memory and Communication Efficient Distributed Stochastic Optimization with Minibatch Prox},'' in \emph{{COLT}}, 2017.

\bibitem{wang2023cocktailsgd}
J.~Wang, Y.~Lu, B.~Yuan, B.~Chen, P.~Liang, C.~De~Sa, C.~Re, and C.~Zhang, ``{CocktailSGD: Fine-Tuning Foundation Models over 500Mbps Networks},'' in \emph{{ICML}}, 2023.

\bibitem{wang2020survey}
M.~Wang, W.~Fu, X.~He, S.~Hao, and X.~Wu, ``{A Survey on Large-Scale Machine Learning},'' \emph{{IEEE Transactions on Knowledge and Data Engineering}}, 2020.

\bibitem{wang2019accelerating}
Z.~Wang, K.~Kara, H.~Zhang, G.~Alonso, O.~Mutlu, and C.~Zhang, ``{Accelerating Generalized Linear Models with MLWeaving: A One-Size-Fits-All System for Any-Precision Learning},'' in \emph{{VLDB}}, 2019.

\bibitem{wortsman2022fi}
M.~Wortsman, S.~Gururangan, S.~Li, A.~Farhadi, L.~Schmidt, M.~Rabbat, and A.~S. Morcos, ``{Lo-fi: Distributed Fine-tuning Without Communication},'' \emph{arXiv:2210.11948}, 2022.

\bibitem{wu2023pim}
Y.~Wu, Z.~Wang, and W.~D. Lu, ``{PIM-GPT: A Hybrid Process-in-Memory Accelerator for Autoregressive Transformers},'' \emph{arXiv:2310.09385}, 2023.

\bibitem{xi2023training}
H.~Xi, C.~Li, J.~Chen, and J.~Zhu, ``{Training Transformers with 4-bit Integers},'' in \emph{{NeurIPS}}, 2023.

\bibitem{xie2017cumf_sgd}
X.~Xie, W.~Tan, L.~L. Fong, and Y.~Liang, ``{CuMF\_SGD: Parallelized Stochastic Gradient Descent for Matrix Factorization on GPUs},'' in \emph{{HPDC}}, 2017.

\bibitem{xu2020compressed}
H.~Xu, C.-Y. Ho, A.~M. Abdelmoniem, A.~Dutta, E.~H. Bergou, K.~Karatsenidis, M.~Canini, and P.~Kalnis, ``{Compressed Communication for Distributed Deep Learning: Survey and Quantitative Evaluation},'' Technical Report, 2020.

\bibitem{xue2024repeat}
F.~Xue, Y.~Fu, W.~Zhou, Z.~Zheng, and Y.~You, ``{To Repeat or Not To Repeat: Insights from Scaling LLM under Token-Crisis},'' in \emph{{NeurIPS}}, 2024.

\bibitem{yu2019parallel}
H.~Yu, S.~Yang, and S.~Zhu, ``{Parallel Restarted SGD with Faster Convergence and Less Communication: Demystifying Why Model Averaging Works for Deep Learning},'' in \emph{{AAAI}}, 2019.

\bibitem{yuksel2024functionally}
{\.I}.~E. Y{\"u}ksel, Y.~C. Tu{\u{g}}rul, A.~Olgun, F.~N. Bostanc{\i}, A.~G. Ya{\u{g}}l{\i}k{\c{c}}{\i}, G.~F. Oliveira, H.~Luo, J.~G{\'o}mez-Luna, M.~Sadrosadati, and O.~Mutlu, ``{Functionally-Complete Boolean Logic in Real DRAM Chips: Experimental Characterization and Analysis},'' in \emph{{HPCA}}, 2024.

\bibitem{Zarif_2023}
N.~Zarif, ``{Offloading Embedding Lookups to Processing-In-Memory for Deep Learning Recommender Models},'' Master's thesis, {University of British Columbia}, 2023.

\bibitem{zhang2014dimmwitted}
C.~Zhang and C.~R{\'e}, ``{DimmWitted: A Study of Main-Memory Statistical Analytics},'' \emph{arXiv:1403.7550}, 2014.

\bibitem{zhang2014top}
D.~Zhang, N.~Jayasena, A.~Lyashevsky, J.~L. Greathouse, L.~Xu, and M.~Ignatowski, ``{TOP-PIM: Throughput-Oriented Programmable Processing in Memory},'' in \emph{{HPDC}}, 2014.

\bibitem{zhang2016parallel}
J.~Zhang, C.~De~Sa, I.~Mitliagkas, and C.~R{\'e}, ``{Parallel SGD: When does averaging help?}'' \emph{arXiv:1606.07365}, 2016.

\bibitem{zhang2019mllib}
Z.~Zhang, J.~Jiang, W.~Wu, C.~Zhang, L.~Yu, and B.~Cui, ``{MLlib*: Fast Training of GLMs Using Spark MLlib},'' in \emph{{ICDE}}, 2019.

\bibitem{zhou2017convergence}
F.~Zhou and G.~Cong, ``{On the Convergence Properties of a K-step Averaging Stochastic Gradient Descent Algorithm for Nonconvex Optimization},'' \emph{arXiv:1708.01012}, 2017.

\bibitem{zhou2020towards}
P.~Zhou, J.~Feng, C.~Ma, C.~Xiong, S.~C.~H. Hoi \emph{et~al.}, ``{Towards Theoretically Understanding Why SGD Generalizes Better Than ADAM in Deep Learning},'' in \emph{{NeurIPS}}, 2020.

\bibitem{zhu2013accelerating}
Q.~Zhu, T.~Graf, H.~E. Sumbul, L.~Pileggi, and F.~Franchetti, ``{Accelerating Sparse Matrix-Matrix Multiplication with 3D-Stacked Logic-in-Memory Hardware},'' in \emph{{HPEC}}, 2013.

\bibitem{zhuo2019graphq}
Y.~Zhuo, C.~Wang, M.~Zhang, R.~Wang, D.~Niu, Y.~Wang, and X.~Qian, ``{GraphQ: Scalable PIM-based Graph Processing},'' in \emph{{MICRO}}, 2019.

\bibitem{zinkevich2010parallelized}
M.~Zinkevich, M.~Weimer, L.~Li, and A.~Smola, ``{Parallelized Stochastic Gradient Descent},'' in \emph{{NeurIPS}}, 2010.

\end{thebibliography}
}

\clearpage
\appendix
\nobalance

\section{Artifact Appendix}

\subsection{Abstract}

\noindent Our artifact~\cite{pimopt-artifact-github,zenodo-artifact} contains the source code and scripts needed to reproduce our results, including all figures in the paper. We provide: 1) source code to preprocess the YFCC100M-HNfc6~\cite{amato2016yfcc100m} and Criteo 1TB Click Logs~\cite{criteo1tb2014} datasets preprocessed by LIBSVM~\cite{LIBSVMData2023}, 2) the source code to perform experiments on the UPMEM PIM System, 3) the source code of the CPU and GPU baseline implementations, and 4) the source code to postprocess and evaluate results. We provide Python scripts and a Jupyter Notebook to analyze and plot the results.

\subsection{Artifact check-list (meta-information)}
\label{appendix:artifact_checklist}
\begin{table}[h]
\centering
{\fontsize{8}{10}\selectfont
\label{tab:appendix-checklist}
\resizebox{\columnwidth}{!}{%
\begin{tabular}{ll}
\textbf{Parameter}        & \textbf{Value}                     \\ \hline
Program &
  \begin{tabular}[c]{@{}l@{}}C programs \\ Python3 scripts/Jupyter Notebook\\ Shell scripts\end{tabular} \\ \hline
Compilation               & \begin{tabular}[c]{@{}l@{}} gcc (Debian 8.3.0-6) 8.3.0 \\ GNU Make 4.2.1\end{tabular} \\ \hline
Run-time environment &
\begin{tabular}[c]{@{}l@{}} Debian GNU/Linux 10 (buster) (UPMEM PIM System) \\ Ubuntu 22.04.1 LTS (CPU Baseline System) \\ Ubuntu 22.04.3 LTS (GPU Baseline System) \\ Python 3.10.6 \\slurm-wlm 21.08.5\\ tmux 2.8+\end{tabular} \\ \hline
Hardware &
  \begin{tabular}[c]{@{}l@{}}2x Intel Xeon Silver 4215 8-core processor @ 2.50GHz,\\ \vspace{0.75em}20×8 GB UPMEM PIM modules (UPMEM PIM System) \\ \vspace{0.75em}2x AMD EPYC 7742 64-core processor @ 2.25GHz (CPU Baseline System) \\ 2x Intel Xeon Gold 5118 12-core processor @ 2.30GHz,\\ 1× NVIDIA A100 (PCIe, 80 GB) (GPU Baseline System)\end{tabular} \\ \hline
Output                    & Data and execution logs in plain text and plots in pdf and png format    \\ \hline
Metrics & \begin{tabular}[c]{@{}l@{}} Runtime, Test Accuracy, AUC Score, \\ Binary Cross Entropy Loss, and Hinge Loss \end{tabular} \\ \hline
Experiment workflow &
  \begin{tabular}[c]{@{}l@{}}Preprocess datasets, perform experiments on UPMEM PIM System, \\ run experiments on CPU Baseline System and GPU baseline,\\postprocess results, and \\ run analysis scripts on the results\end{tabular} \\ \hline
Disk space requirement    & $\approx$ 12TB                 \\ \hline
Workflow preparation time & \begin{tabular}[c]{@{}l@{}}   $\approx$ 3 days to preprocess YFCC100-HNfc6 dataset\\ $\approx$ 20 hours to preprocess Criteo 1TB Click Logs dataset  \end{tabular} \\ \hline
Experiment completion time &
  \begin{tabular}[c]{@{}l@{}}   $\approx$ 2 days to perform experiments using the UPMEM PIM System \\ on the YFCC100M-HNfc6 dataset \\ $\approx$ 1 week to perform experiments using the UPMEM PIM System \\ on the Criteo 1TB Click Logs dataset \\ $\approx$ 16 hours to perform experiments using the CPU Baseline System \\ on the YFCC100M-HNfc6 dataset \\ $\approx$ 8 hours to perform experiments using the CPU Baseline System \\ on the Criteo 1TB Click Logs dataset \\ $\approx$ 1 hour to perform experiments using the GPU Baseline System \\ on the YFCC100M-HNfc6 dataset \\ $\approx$ 12 hours to postprocess CPU and GPU Baseline System results  \\ $\approx$ 2 days to postprocess UPMEM PIM System results  \\ $\approx$ 1 hour to aggregate results, and reproduce plots\end{tabular} \\ \hline
Publicly available?       &  \begin{tabular}[c]{@{}l@{}} Zenodo (\url{https://doi.org/10.5281/zenodo.12747665})       \\ Github (\url{https://github.com/CMU-SAFARI/PIM-Opt}) \end{tabular}       \\\hline
Code licenses             & MIT                                \\ \hline
\end{tabular}
}
}
\end{table}

\subsection{Description}

\noindent \subsubsection{How to access}
\label{appendix:how_to_access}

The artifact is available on Zenodo with DOI \url{https://doi.org/10.5281/zenodo.12747665}. The live Github repository is at \url{https://github.com/CMU-SAFARI/PIM-Opt}.

\noindent \subsubsection{Hardware dependencies}
\label{appendix:hardware_dependencies}
Our hardware dependencies are listed in Table~\ref{table:dpu_system-configuration}. For preprocessing of the datasets (\secref{appendix:preprocessing_datasets}) and postprocessing of the results (\secref{appendix:postprocessing_results}), the same hardware configuration as the CPU Baseline System is used.

\noindent \subsubsection{Software dependencies} 
\label{appendix:software_dependencies}

\begin{itemize}
    \item \texttt{gcc (Debian 8.3.0-6) 8.3.0, GNU Make 4.2.1}
    \item \texttt{UPMEM SDK}, version 2023.2.0~\cite{upmem_sdk}
    \item \texttt{tar (GNU tar) 1.34}
    \item \texttt{Zip 3.0}
    \item \texttt{Python 3.10.6}
    \item \texttt{pip} packages \texttt{pandas, numpy, scipy, scikit-learn, matplotlib, seaborn, torch, coloredlogs}
    \item \texttt{slurm-wlm 21.08.5}
    \item \texttt{tmux 2.8+}
    \item \texttt{CUDA 11.7}
\end{itemize}

\noindent \subsubsection{Datasets}
\noindent In this paper, we use two large-scale datasets: \begin{itemize}
    \item YFCC100M-HNfc6~\cite{amato2016yfcc100m} can be requested at \url{http://www.deepfeatures.org/index.html}. For preprocessing, one needs to download the file {\texttt{yfcc100m\_autotags.bz2}} from the original YFCC100M dataset~\cite{yfcc100m}, which can be requested at \url{https://www.multimediacommons.org}.
    \item Criteo 1TB Click Logs~\cite{criteo1tb2014} preprocessed by LIBSVM~\cite{LIBSVMData2023} can be accessed by running:
        \shellcmd{wget -t inf https://www.csie.ntu.edu.tw/\mytexttilde cjlin/\ libsvmtools/datasets/binary/criteo\_tb.svm.tar.xz}
        \shellcmd{tar -xJvf criteo\_tb.svm.tar.xz}
\end{itemize}

\subsection{Installation}
\label{appendix:installation}

\noindent To reproduce our results, no extra installation steps are required besides installing the dependencies described in \secref{appendix:software_dependencies}. We recommend using a terminal multiplexer (e.g., \texttt{tmux}) to ensure that experiments are completed without interruption.

\subsection{Experiment workflow}
\label{appendix:exp_workflow}

\om{We describe the steps and commands to reproduce our results, including all figures in the paper, in this section. Note that we assume the use of \texttt{slurm} workload manager on a cluster. Readers with other workload managers should modify the scripts to fit their own environment.}

\noindent \subsubsection{Preprocessing Datasets}
\label{appendix:preprocessing_datasets}
\noindent The preprocessing of the datasets YFCC100M-HNfc6 and Criteo is initialized by running the commands:
            \shellcmd{cd preprocessing}
            \shellcmd{DATA\_ROOT=<path-to-data>}
            \shellcmd{PARTITION=<name-of-slurm-partition>}
            \shellcmd{NODE=<name-of-slurm-node>}
            \shellcmd{./run\_preprocessing.sh \$\{DATA\_ROOT\} \$\{PARTITION\} \$\{NODE\} \&}

\noindent \subsubsection{UPMEM PIM System Experiments}
\label{appendix:upmem_pim_system_experiments}
\noindent To perform the experiments on the UPMEM PIM system, readers can run the command:
            \shellcmd{cd upmem\_ml\_coding/UPMEM}
            \shellcmd{DATA\_ROOT=<path-to-data>}
            \shellcmd{./run\_upmem\_experiments.sh  \$\{DATA\_ROOT\} \&}

\noindent \subsubsection{CPU and GPU Baseline Experiments}
\label{appendix:cpu_gpu_baseline_experiments}
The baseline experiments are launched by running the commands:
            \shellcmd{cd baseline}
            \shellcmd{DATA\_ROOT=<path-to-data>}
            \shellcmd{PARTITION=<name-of-slurm-partition>}
            \shellcmd{NODE\_CPU=<name-of-slurm-cpu\_node>}
            \shellcmd{NODE\_GPU=<name-of-slurm-gpu\_node>}
            \shellcmd{./run\_baseline\_experiments.sh \$\{DATA\_ROOT\} \$\{PARTITION\}} 
            \shellcommand{\quad\$\{NODE\_CPU\} \$\{NODE\_GPU\} \&}

\noindent \subsubsection{Postprocessing Results}
\label{appendix:postprocessing_results}
\noindent Before continuing, the experiments in \secref{appendix:upmem_pim_system_experiments} and \secref{appendix:cpu_gpu_baseline_experiments} must be completed. To continue with the postprocessing of the UPMEM \srm{PIM system} results, i.e., computing metrics such as AUC Score, place the UPMEM PIM system results into the directory \texttt{/results}. Next, please run the commands:
            \shellcmd{cd postprocessing}
            \shellcmd{DATA\_ROOT=<path-to-data>}
            \shellcmd{PARTITION=<name-of-slurm-partition>}
            \shellcmd{NODE=<name-of-slurm-node>}
            \shellcmd{./run\_postprocessing\_Criteo.sh  \$\{DATA\_ROOT\} \$\{PARTITION\}}
            \shellcommand{\quad\$\{NODE\} \&}

\noindent \subsubsection{Reproducing Figures}
\label{appendix:reproducing_figures}

\noindent Please navigate to the directory \texttt{/paper\_plots}, open the Jupyter Notebook \texttt{paper\_plots.ipynb}, and select \texttt{Run All} or if you prefer, you can click through the Jupyter Notebook cell by cell. The generated figures can be viewed at \texttt{/paper\_plots/output} in \texttt{pdf} and \texttt{png} format.

\subsection{Evaluation and expected results}
\noindent Running the experiments described in (\secref{appendix:exp_workflow}) is sufficient to reproduce all of our results (Fig.~\ref{fig:motivational_figure}, Fig.~\ref{fig:yfcc_perf_breakdown}, Fig.~\ref{fig:yfcc_perf_comparison}, Fig.~\ref{fig:yfcc_batch_size_sweep_compact}, Fig.~\ref{fig:yfcc_weak_scaling}, Fig.~\ref{fig:yfcc_strong_scaling}, Fig.~\ref{fig:criteo_perf_breakdown}, Fig.~\ref{fig:criteo_perf_comparison}, Fig.~\ref{fig:criteo_batch_size_sweep_compact}, Fig.~\ref{fig:criteo_weak_scaling}, and Fig.~\ref{fig:criteo_strong_scaling}).

\end{document}